\documentclass[acmsmall]{acmart}

\usepackage{longtable}
\usepackage{multirow}
\usepackage{lineno}
\usepackage{soul}
\usepackage{titlesec}
\titleclass{\subsubsubsection}{straight}[\subsection]

\newcounter{subsubsubsection}[subsubsection]
\renewcommand\thesubsubsubsection{\thesubsubsection.\arabic{subsubsubsection}}
\titleformat{\subsubsubsection}
  {\normalfont\normalsize\bfseries}{\thesubsubsubsection}{1em}{}
\titlespacing*{\subsubsubsection}
{0pt}{3.25ex plus 1ex minus .2ex}{1.5ex plus .2ex}

\makeatletter
\renewcommand\paragraph{\@startsection{paragraph}{5}{\z@}%
  {3.25ex \@plus1ex \@minus.2ex}%
  {-1em}%
  {\normalfont\normalsize\bfseries}}
\renewcommand\subparagraph{\@startsection{subparagraph}{6}{\parindent}%
  {3.25ex \@plus1ex \@minus .2ex}%
  {-1em}%
  {\normalfont\normalsize\bfseries}}
\def\toclevel@subsubsubsection{4}
\def\toclevel@paragraph{5}
\def\toclevel@paragraph{6}
\def\l@subsubsubsection{\@dottedtocline{4}{7em}{4em}}
\def\l@paragraph{\@dottedtocline{5}{10em}{5em}}
\def\l@subparagraph{\@dottedtocline{6}{14em}{6em}}
\makeatother

\AtBeginDocument{%
  \providecommand\BibTeX{{%
    \normalfont B\kern-0.5em{\scshape i\kern-0.25em b}\kern-0.8em\TeX}}}

\acmJournal{CSUR}
\begin{document}

\title{Computing Server Power Modeling in a Data Center: Survey, Taxonomy and Performance Evaluation}

\author{Leila Ismail}
\authornote{Corresponding Author}
\author{Huned Materwala}
\email{emails: {leila, huned.m}@uaeu.ac.ae}
\affiliation{%
  \institution{Department of Computer Science and Software Engineering, College of Information Technology, United Arab Emirates University}
  \streetaddress{P.O. Box 15551}
  \city{Al Ain}
  \country{United Arab Emirates}
}


\begin{abstract}
 Data centers are large scale, energy-hungry infrastructure serving the increasing computational demands as the world is becoming more connected in smart cities.  The emergence of advanced technologies such as cloud-based services, internet of things (IoT) and big data analytics has augmented the growth of global data centers, leading to high energy consumption.  This upsurge in energy consumption of the data centers not only incurs the issue of surging high cost (operational and maintenance) but also has an adverse effect on the environment.  Dynamic power management in a data center environment requires the cognizance of the correlation between the system and hardware level performance counters and the power consumption.  Power consumption modeling exhibits this correlation and is crucial in designing energy-efficient optimization strategies based on resource utilization.  Several works in power modeling are proposed and used in the literature.  However, these power models have been evaluated using different benchmarking applications, power measurement techniques and error calculation formula on different machines. In this work, we present a taxonomy and evaluation of 24 software-based power models using a unified environment, benchmarking applications, power measurement technique and error formula, with the aim of achieving an objective comparison.  We use different servers architectures to assess the impact of heterogeneity on the models' comparison.  The performance analysis of these models is elaborated in the paper.
\end{abstract}

%\begin{CCSXML}
%<ccs2012>
%  <concept_id>10010520.10010553.10010562</concept_id>
%  <concept_desc>Computer systems organization~Embedded systems</concept_desc>
%  <concept_significance>500</concept_significance>
%  <concept_id>10010520.10010575.10010755</concept_id>
%  <concept_desc>Computer systems organization~Redundancy</concept_desc>
%  <concept_significance>300</concept_significance>
%  <concept_id>10010520.10010553.10010554</concept_id>
%  <concept_desc>Computer systems organization~Robotics</concept_desc>
%  <concept_significance>100</concept_significance>
%  <concept_id>10003033.10003083.10003095</concept_id>
%  <concept_desc>Networks~Network reliability</concept_desc>
%  <concept_significance>100</concept_significance>
%</ccs2012>
%\end{CCSXML}
%
%\ccsdesc[500]{Computer systems organization~Embedded systems}


\maketitle

\section{Introduction}
Data centers are substantial computing facilities serving as a back-end infrastructure for enabling globally competitive innovations and contributing to the socio-economic development \cite{Buyya2010,Dayarathna2016}.  There is rapid growth to data centers comprising of thousands of computing nodes due to the emergence of smart cities and consequently the need of paradigms such as Cloud Computing \cite{Mell2011}, IoT \cite{Buyya:2016:ITP:3050877} and Big Data Analytics \cite{IntelITCenter2012}.  This continuous storage and computing needs lead technical firms like Microsoft and Google to expand their data center infrastructures as large as a football field able to host thousands of nodes\cite{Buyya2013}.  The data center services market is projected to grow at a compound annual growth rate (CAGR) of 13.69\% over the forecast period of 2018-2023 \cite{GlobalDa23:online}.  A data center, however, has a massive energy consumption engendering various economic problems and environmental hazards.

Energy consumption of data centers is becoming an important issue in an enterprise environment and has gained significant importance in recent years.  A typical data center may consume energy equivalent to that of 25,000 households \cite{Buyya2013}.  According to the report by National Resources Defense Council (NRDC) in the USA, the data centers in 2013 consumed 91 billion kWh of energy, comparable to 34 large power plants (coal fired) \cite{Americas54:online}.  This energy consumption of the data center is anticipated to reach around 140 billion kWh by 2020, equivalent to the annual output of 50 power plants, incurring the cost of \$13 billion in electricity to the American business.  Furthermore, a typical data center's energy cost increases by 100\% every five years \cite{Buyya2013}.  The carbon emissions caused by data centers in 2005 in the USA was as much as that of a mid-sized nation like Argentina \cite{Mathew2012}.  It is expected that by 2020 the annual carbon emission of the data centers will reach 100 million metric tons \cite{Americas54:online}.

The data center's power consumption comprises of \cite{greenberg2008cost,pelley2009understanding,zheng2014joint}: 1) the power consumed by the data center's IT equipment such as computing servers and storage (56\%), 2) the power consumed by the infrastructure facilities such as the cooling systems (30\%), the power distribution/conditioning systems (8\%), and the lighting (1\%), and 3) the power consumed by the network (5\%) \cite{pelley2009understanding,zheng2014joint}.  To reduce the data center energy consumption, different methods have been introduced in the literature, such as deploying energy-efficient algorithms, modifying the hardware components architecture \cite{David2010}, designing measures for efficient air handling \cite{Steve2006} and cooling \cite{GRID}, and using efficient options for the power supply.  These methods require modeling of the relationship between systems' power consumption (considered as dependent variable) and performance counters (considered as independent variable) \cite{Dwang}.  Consequently, data center operators  need an accurate power  model for designing an energy efficient system \cite{Rivoire2008,DBLP:conf/e2dc/BergeCKOPPVW12}, managing a center power consumption \cite{Kilper2011} and using energy-aware scheduling for optimization \cite{Xu2013a}.  These power models proposed in the literature are classified as 1) hardware-based models that use fan speed, voltage, current, capacitance, resistance, and motherboard components as the independent variables, and 2) software-based models that target either individual subsystems of a server, such as CPU, memory, disk and network, or a virtual machine, or a full-system (a computing server) \cite{mobius2013power}.  We use the terminology computing server for a full-system in the remainder of the paper.  The hardware-based power models require sensors to measure different variables on a server's hardware.  This adds in extra hardware and energy consumption costs incurred by these sensors attached to thousands of servers in a data center.  However, the software-based models do not require external sensors to get the values of the model variables adding no extra cost.  Therefore, in this paper, we focus on the software-based power models.  The software models use performance metrics provided by the operating system that we call System\_Performance\_Metrics (S\_PM) based models, or performance monitoring counters provided by the hardware subsystems of a server that we call System\_Performance\_Counters (S\_PC) based, or a combination of system performance metrics and counters that we call System\_Performance\_Metrics\_Counters (S\_PMC) based.  The metrics provided by the operating system indicate the utilization level of a system (CPU utilization, memory utilization, disk I/O rate and network I/O rate), whereas the counters provided by the hardware indicate the performance of the different server's subsystems, such as number of cache misses \cite{CPUcache77:online}, number of branch instructions \cite{Branchco63:online}, and number of interrupts  \cite{Interrup56:online}.  In this work, we evaluate the performance of software-based computing server's power consumption models that have been proposed in the literature.  However, we are unaware of any objective comparison of these models using a unified experimental setup for a diverse set of applications.  This paper focuses to address this void.

In this study, we present a taxonomy and comparative evaluation of software-based power models. We evaluate their performance in terms of standard error of estimation.  This is in a unified environment and experimental setup.  In this evaluation, we make use of four different tools for models formulation and validation and five different applications for models testing.  The key research contributions of this work are as follows.
\begin{itemize}
\item We classify the work on power modeling into software and hardware-based models, and present a taxonomy of different software-based power models in the data center's power consumption modeling literature.  We discuss the temporal evolution of the various models and capture the assumptions conducting to the development of a given model in a certain period of time.
\item We evaluate the performance of 24 different software-based power models in terms of standard error of estimation using a diverse set of benchmarking applications in a unified experimental setup.  The experiments show that the support vector machine (SVM) power model has the least standard error of estimation.
\item The portability of the relationship between a server's power consumption and the user and system performance counters is also verified on different server architectures in our experimental testbed.
\end{itemize}

To our knowledge, this is the first work to classify software-based power models in the literature and evaluate their performance in a unified environment and setup.

The rest of the paper is organized as follows.  Section II synthesizes a taxonomy of the works on software-based power models.  The experimental setup, experiments and the performance evaluation in terms of standard error of estimation for the studied power models are described in Section III.  Section IV overviews the related works.  The paper is concluded with the lessons learned and possible future research directions in Section V.

\section{Software-based Power Models}
We first present the taxonomy (Figure \ref{fig:taxonomy}) and limitations (Table \ref{table:one}) of state-of-the-art software-based power models (Sub-section 2.1).  This allows to capture which were the assumptions conducting to the development of a given model in a certain period of time and unravel on one hand the technological development of servers and, on the other hand, the corresponding improvements in the precision of the models.  In sub-section 2.2, we recall the experimental setups used for the evaluation of these models in the literature along with their precision (Table 2) and describe the workflow of power model development system that is used to build the studied power models. 

\subsection{Taxonomy of Software-based Power Models}
\begin{figure}
\centerline{\includegraphics[width=\linewidth]{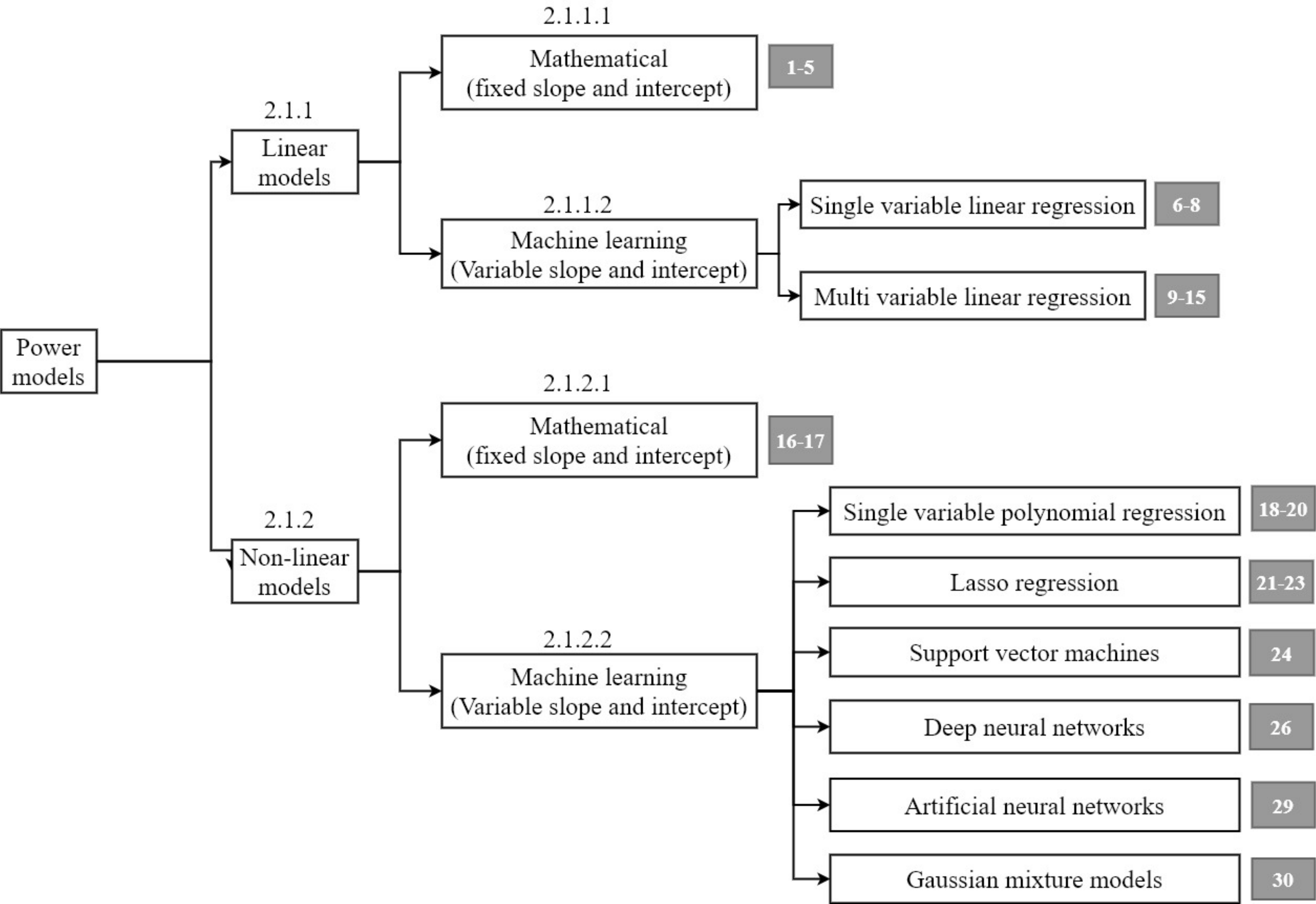}}
\caption{Taxonomy of the power models.  The numbers above the classified categories denote the sub-subsection and sub-sub-subsection, and the gray box besides each category represents the corresponding equation numbers of the power models.}
\label{fig:taxonomy}
\end{figure}

We present a classification and temporal evolution of the software-based power models in the literature.   We classify these models into two primary categories: 1) linear and 2) non-linear.  We further classify the models in these categories into 1) mathematical formula based on fixed slope and intercept and 2) machine learning based on variable slope and intercept.  The former is based on a server's idle and full load power consumption values.  It does not consider the implication of the spatial distribution of power consumption for the S\_PM and S\_PC values which lie between the idle and the full load states of the server.  Whereas, the machine learning models are developed based on the distribution of the power consumption of the S\_PM and S\_PC utilization values.  Figure \ref{fig:taxonomy} shows our taxonomy of power models in the literature.

\graphicspath{{./Images/}}

\subsubsection{Linear Power Models}

\subsubsubsection{Mathematical Formula: Single Variable Linear With Fixed Slope and Intercept (SVLF)}
In the 2000s, with the research attention drawn towards the energy efficiency of the data centers due to the onset of large-scale web services and the underlying massive parallel computing infrastructure, studies were conducted to analyze the computing servers' hardware power consumption breakdown \cite{Bohrer2002,Fan2007}.  \cite{Bohrer2002} studied the power consumption of web servers using real workloads derived from the logs of three production websites (1998 Winter Olympics, a financial services company, and Information Resource Caching project affiliated with the National Laboratory for Applied Network Research).  Results from this study stated that the server's power consumption is highly dominated by its CPU utilization in a linear manner, influencing many works, that come after, on server power consumption modeling.  In 2006, \cite{Heath2006} formulated a linear model as stated in Equation \ref{one} to calculate power consumption (P) of a server with CPU utilization ($u_{cpu}$) as the independent variable (S\_PM-based).  The slope of the linear model is the difference between the power consumption when the server is at full load ($P_{MAX}$) and the power consumption when the server is idle ($P_{MIN}$),  and the intercept of the model being  $P_{MIN}$. However, the model was not evaluated for its accuracy by the authors.  Later in 2007, \cite{Fan2007} studied the power usage of thousands of servers for workloads taken from different classes of web services such as Websearch, Webmail, and Mapreduce for over a period of approximately six months and confirmed the results obtained by \cite{Bohrer2002}.
\begin{equation}
\label{one}
P = (P_{MAX}\textrm{ - }P_{MIN})\times u_{cpu} + P_{MIN}
\end{equation}

The model in Equation \ref{one} was later evaluated by \cite{Fan2007} and \cite{CHEUNG2018329} for its accuracy.  The model has been used by various works in the literature for electricity consumption cost prediction in a heterogeneous server environment \cite{Qureshi2009}, for energy-aware resource management technique \cite{Gmach2009}, and for energy-efficient cloud computing \cite{Lee2012,Raycroft2014,Dai2016,Sharma2016,7997707}.  The model in Equation \ref{one} was then presented by \cite{Buyya2010} using the ratio $\frac{P_{MIN}}{P_{MAX}}$ denoted by $k$ as stated in Equation \ref{two}.  The slope $P_{MAX} - P_{MIN}$ and the intercept $P_{MIN}$ of Equation \ref{one} are then represented as $[1-k]\times P_{MAX}$ and $k\times P_{MAX}$ respectively in Equation \ref{two}.  Substituting the value of $k$ in Equation \ref{two} yields to Equation \ref{one}, making them identical.  The model has been used by various works on energy efficient cloud resource management in the literature \cite{Hongyou2013,Bagheri2014,Tang2015,Patel2015}.
\begin{equation}
\label{two}
P = ([1-k]\times P_{MAX}\times u_{cpu}) + (k\times P_{MAX})
\end{equation}

Later in the year 2007, \cite{Fan2007} showed that an idle server almost consumes 70\% of its peak power consumption (i.e. \textrm{k} = 0.7).  Based on this study, \cite{Beloglazov2012a} stated a S\_PM-based power model as in Equation \ref{three} derived from Equation \ref{two} by substituting the value of $k$ as 0.7.  The model has been used in the literature for data center energy efficiency \cite{Sinha2011,Han2016}.
\begin{equation}
\label{three}
P = P_{MAX}\times (0.7 + 0.3\times u_{cpu})
\end{equation}

In 2011, with the advancement of cloud computing technology wherein computing infrastructure and applications are provided as services to end users under pay-per-use model, \cite{Calheiros:2011:CTM:1951445.1951450} developed CloudSim software for modeling and simulation of cloud computing environments and evaluation of resource provisioning algorithms.  The power model used in the simulation software to predict the power consumption for executing the cloud applications is based on the method of linear interpolation \cite{alma9928725064401631} using CPU utilization as the independent variable (S\_PM-based) as stated in Equation \ref{four}.  Linear interpolation assumes a piece-wise linear relation between each interval of CPU utilization values and corresponding power consumption values instead of assuming a single linear function between idle and full load utilization.  The model was later adopted by various works using CloudSim to evaluate the energy efficiency of the cloud resource allocation algorithms \cite{Beloglazov2012,Kord2013,Farahnakian2014,Chowdhury2015,Arianyan2015,Rahmanian2017,Wen2017,Kejing2017}.
\begin{equation}
\label{four}
P = P_1 + (\Delta\times (\frac{u_{cpu}-{u_{1_{cpu}}}}{10})\times 100)
\end{equation}
where \textit{${P_1}$} and \textit{${P_2}$} are the power values corresponding to the CPU utilization \textit{${u_1}$} and \textit{${u_2}$} respectively ($u_1$ $\leq$ u $\leq$ $u_2$) and \textit{$\Delta$} is the slope of the line between points $(u_1, P_1)$ and $(u_2, P_2)$.

In 2013, with the accelerating adoption of cloud computing serving applications of type I/O and web-based interaction, a significant amount of network is then used due to the virtualization of computing servers.  Therefore, \cite{Jin2013} studied the energy consumption of servers in a cloud computing environment for energy efficiency for network-intensive benchmarks and developed the relationship between the power consumed by an application running on a server and the application's throughput as stated in Equation \ref{five}.  The power model is S\_PM-based, using the application's throughput as the independent variable.
\begin{equation}
\label{five}
P = ([P_{MAX}\textrm{ - }P_{MIN}]\times \frac{Throughput}{Throughput_{Max}}) + P_{MIN}
\end{equation}
where the throughput is application specific and is defined as the amount of load executed per unit of time.  For instance, the throughput of a network-intensive HTTP application is known as the request rate defined by the number of requests processed per second.

\subsubsubsection{Machine Learning Linear With Variable Slope and Intercept (MLLV)}

\begin{itemize}
\item \textit{Single Variable Linear Regression (SVLR):}
In 2008, \cite{Raghavendra2008} proposed a power management solution for data center based on a predicted power consumption value using a fitted-line regression model as stated in Equation \ref{eight}.  The authors experimentally collected the power consumption values of a server while running workloads at different CPU utilization values to develop this S\_PM-based model.  The linear regression \cite{Mathematics} model develops a linear relationship between the power consumption and the CPU utilization by calculating an intercept and a slope for the linear line.  The model was later used by the works on energy-aware scheduling in a data center \cite{Gong2008,Berral2010,Berral2011} and the works on power consumption modeling \cite{Pedram2010,Zhang2013}.
\begin{equation}
\label{eight}
P = \alpha + \beta u_{cpu}
\end{equation}
where \textit{$\alpha$} and \textit{$\beta$} are the intercept and slope of the regression line whose values are calibrated for each server type experimentally such that the squared error is the minimum.

In 2010, \cite{Wang2010} used the same linear regression model stated in Equation \ref{eight} for power and cooling management in the data centers.  The authors fixed the value of intercept at server's idle power consumption as stated in Equation \ref{nine}, instead of calibrating the value of intercept experimentally (Equation \ref{eight}).
\begin{equation}
\label{nine}
P = P_{MIN} + \beta u_{cpu}
\end{equation}

In 2010,\cite{Koller2010} conducted experiments on servers using web-transactions, HPC, and I/O-intensive workloads and found that the linear regression power model based on the CPU utilization only (Equation \ref{eight}) predicts the power consumption with a large error.  Consequently, the authors proposed a S\_PM-based power model to establish a linear relationship between the power consumption of an application and the application's throughput to predict the power of heterogeneous applications as stated in Equation \ref{ten}.
\begin{equation}
\label{ten}
P = \alpha + \beta(Throughput)
\end{equation}
where, throughput is application specific and defined as the number of requests executed per second.

\item \textit{Multi Variable Linear Regression (MVLR):}
In 2006, \cite{Economou2006} conducted experiments on a blade server to study its metric-level power consumption (CPU, memory, disk, and network) using different benchmarks such as SPECcpu2000 integer, SPECcpu2000 floating point, SPECjbb2000, SPECweb2005, the streams, and the matrix multiplication.  Based on these results, the authors stated that the memory power consumption is likely to be equally important, if not more, as that of the CPU.  Moreover, the power consumed by the disk and the network I/O peripherals can not be neglected.  Consequently, the authors proposed a power consumption model based on multi linear regression as stated in Equation \ref{fourteen}, with CPU utilization (\textit{$u_{cpu}$}), memory utilization (\textit{$u_{mem}$}), disk I/O rate (\textit{$u_{disk}$}) and network I/O rate (\textit{$u_{net}$}) as the independent variables (S\_PM-based).  The model was later used by \cite{Smith2010} while profiling power usage in cloud computing environment.  In 2014, \cite{Alan2014} proposed multi regression model as stated in Equation \ref{fifteen} using the performance metrics similar to that in Equation \ref{fourteen}.  However, the model used the server's idle power consumption as the intercept of the model instead of calculating it based on the regression fitted-hyperplane.
\begin{equation}
\label{fourteen}
P = \alpha + \beta_1u_{cpu} + \beta_2u_{mem} + \beta_3u_{disk} + \beta_4u_{net}
\end{equation}
\begin{equation}
\label{fifteen}
P = P_{MIN} + \beta_1u_{cpu} + \beta_2u_{mem} + \beta_3u_{disk} + \beta_4u_{net}
\end{equation}

where \textit{$\alpha$}, \textit{$\beta_1$}, \textit{$\beta_2$}, \textit{$\beta_3$}, and \textit{$\beta_4$} are the model parameters calibrated for each server type experimentally such that the squared error of estimation is the minimum.

In 2010, \cite{Kansal2010} examined the power consumption of a server and found that the most power consuming components of a server are the CPU, the memory and the disk.  Based on these results, the authors proposed a S\_PM-based multi regression power model as stated in Equation \ref{sixteen}.  The model was later used by \cite{Li2012}.
\begin{equation}
\label{sixteen}
P = \alpha + \beta_1u_{cpu} + \beta_2u_{mem} + \beta_3u_{disk}
\end{equation}

Later in 2010, \cite{husain} conducted experiments and found that the power consumption of a server is a linear function of the CPU load, memory utilization, disk operations per second and instruction cache.  Based on these results, the authors proposed a model based on  S\_PMC as in Equation \ref{seventeen}.  In the same year, \cite{DaCosta2010} conducted experiments to study the energy consumed by different applications.  The authors used synthetic workload to stress the server and simultaneously measured the power consumption and 165 server performance indexes.  Nine significant S\_PMC, such as the square of CPU utilization, context switches, cache references, square of cache misses, disk read/write per second, number of TLB interrupts, RES interrupts, NMI interrupts and LOC interrupts corresponding to power consumption are then used to develop a fine-grained power model as stated in Equation \ref{eighteen}.
\begin{equation}
\label{seventeen}
P = \alpha + \beta_1u_{cpu} + \beta_2u_{mem} + \beta_3u_{disk}+ \beta_4cache
\end{equation}
\begin{equation}
\label{eighteen}
P = P_{MIN} + \sum_{n=1}^{9} \beta_nC_n
\end{equation}

where \textit{$C_n$} are the performance counters.

In 2012, \cite{witkowski2013practical} studied different hardware and software power measurement solutions for HPC applications. Based on the analysis presented, the hardware solutions are expensive compared to the software ones. The authors proposed S\_PMC based model for power prediction as stated in Equation \ref{temperatureMLR2} using LLC load misses, LLC loads, LLC store misses, LLC stores, branch misses, branches, cache misses, cache references, context switches, cycle, dTLB load misses, dTLB loads, dTLB store misses, dTLB stores, iTLB load misses, iTLB loads, instructions, major and minor faults, page faults, CPU utilization, number of bus transactions, and DRAM access as the independent variables.  In addition to these variables, the model also used CPU temperature and frequency.  The authors developed different models using only the highest CPU frequency and using all the CPU frequencies.  The authors studied the correlation of S\_PMC with the power consumption and used all the variables that increased the prediction accuracy, starting from the one which has the highest correlation.

\begin{equation}
\label{temperatureMLR2}
P = \beta_0 + \sum_{n=1}^{33} \beta_nC_n
\end{equation}

Later in 2013, \cite{jarus2014runtime} studied the power consumption of HPC servers using a set of real-life applications.  The authors proposed the use of clustering approach to group the applications having similar power characteristics.  Each group of applications have a different power model.  The selection of power model is done automatically using decision tree.  The authors proposed S\_PC-based power model as stated in Equation \ref{temperatureMLR} using 24 performance counters namely: LLC load misses, LLC loads, LLC stores, LLC store misses, branches, cache misses, cache references, context switches, cycles, dTLB load misses, dTLB loads, dTLB stores, dTLB store misses, iTLB loads, iTLB load misses, instructions, major faults, minor faults, page faults, DRAM access 1 and 2, and number of bus transactions.  In addition to these counters, the model also uses CPU temperature.

\begin{equation}
\label{temperatureMLR}
P = \alpha + \sum_{n=1}^{24} \beta_nC_n
\end{equation}
\end{itemize}

\subsubsection{Non-Linear Models}
\subsubsubsection{Mathematical Formula: Single Variable Non-Linear With Fixed Slope and Intercept (SVNLF)}
In 2007, while studying the power usage characteristics of servers, \cite{Fan2007} confirming that the power consumption of a server is highly dominated by the server's CPU utilization, the authors presented, in addition to the linear power model stated previously in Equation \ref{one}, an empirical non-linear power model.  The authors performed experiments on thousands of heterogeneous servers and found that S\_PM-based non-linear power model, stated in Equation \ref{six}, fits the power consumption curve of the server better than the linear model (Equation \ref{one}).  This non-linear model was later used by various works on power consumption modeling \cite{Qureshi2009,Sun2015PowerEfficientPF}.
\begin{equation}
\label{six}
P = ([P_{MAX} \textrm{ - } P_{MIN}]\times [2u_{cpu} \textrm{ - } u_{cpu}^r]) + P_{MIN}
\end{equation}
where \textit{r} is the calibration parameter whose value is obtained experimentally for each server type such that it minimizes the squared error of estimation.

Later in 2007, considering the increasing concern of power consumption in streaming-media servers, \cite{Lien2007} studied the power behavior of the media servers.  The authors performed experiments and observed the power consumed by different media servers using streaming media workloads and found that the power consumption of the servers is based on the idle power and the full load power, and the power consumption is related to the CPU utilization of the server.  The results of the experiments showed that the power consumption of the media servers increases non-linearly when the CPU utilization of the server increases from 0\% to 100\%.  Based on these observations, \cite{Lien2007} considered that the power consumption of a streaming-media server as an exponential function of the server's CPU utilization and proposed a S\_PM-based power model as stated in Equation \ref{seven}.  The model was later used by \cite{Tang2011} for energy efficient resource management in cloud service data centers.
\begin{equation}
\label{seven}
P = ([P_{MAX} \textrm{ - } P_{MIN}]\times [\alpha u_{cpu}^{\beta}]) + P_{MIN}
\end{equation}
where \textit{$\alpha$} and \textit{$\beta$} are the model parameters calibrated for each server type experimentally such that the squared error of estimation is the minimum.

\subsubsubsection{Machine Learning Non-Linear With Variable Slope and Intercept (MLNLV)}

\begin{itemize}
\item \textit{Single Variable Polynomial Regression (SVPR):}
In 2012, \cite{Janacek2012} extended the linear regression power model based on CPU utilization only (Equation \ref{eight}) to quadratic model as stated in Equation \ref{eleven} based on the fact that the power consumption of a server is proportional to the square of the CPU frequency.  This relationship between a server's power consumption and the CPU frequency is due to Dynamic Voltage and Frequency Scaling (DVFS) capability exhibited by a server with its advancement.  DVFS is a technique that dynamically adjusts the voltage and frequency of a server's CPU to optimize resource allocation and maximize power savings when the resources are not required \cite{le2010dynamic}.  In the same year, \cite{Guazzone2012} presented a power model having a polynomial of degree 'r' as stated in Equation \ref{twelve} instead of using the quadratic polynomial as in Equation \ref{eleven} to avoid over-fitting of the model.  In 2013, \cite{Zhang2013} performed experiments on seven heterogeneous servers using the SPECpower benchmark \cite{SPECpowe70:online} to analyze the accuracy of the linear regression model based on the CPU utilization (Equation \ref{eight}).  The results showed that not all the servers hold the linear relationship between power consumption and CPU utilization.  Based on this finding, the authors extended the linear model to higher degree polynomial models of degree two and three as stated in Equations \ref{eleven} and \ref{thirteen} respectively and found that the polynomial models are more accurate than the linear model.  The models in Equations \ref{eleven}-\ref{thirteen} are S\_PM-based using CPU utilization as the independent variable.  In 2016, \cite{canuto2016methodology} used polynomial regression to capture the non-linear behavior between S\_PM independent variables and server power consumption.

\begin{equation}
\label{eleven}
P = \alpha + \beta_1u_{cpu} + \beta_2u_{cpu}^2
\end{equation}
\begin{equation}
\label{twelve}
P = \alpha + \beta_1u_{cpu} + \beta_2u_{cpu}^r
\end{equation}
\begin{equation}
\label{thirteen}
P = \alpha + \beta_1u_{cpu} + \beta_2u_{cpu}^2 + \beta_3u_{cpu}^3
\end{equation}

where \textit{$\alpha$}, \textit{$\beta_1$}, \textit{$\beta_2$}, \textit{$\beta_3$}, and \textit{r} are the model parameters calibrated for each server type experimentally such that the squared error of estimation is the minimum.

\item \textit{Lasso Regression (LR):}
In 2011, \cite{mccullough2011evaluating} studied the power consumption of servers corresponding to the CPU and memory utilization values.  The authors found that the CPU and memory variables are independent of each other and consequently using a linear regression model based on these variables might not produce accurate results.  The authors proposed the use of polynomial regression of order 3 with CPU and memory utilization values as the independent variables.  Consequently, the S\_PM-based model has linear, quadratic, and cubic functions of the CPU and memory utilization values (CPU, $CPU^2$, $CPU^3$, mem, $mem^2$, $mem^3$).  To reduce the number of estimators, the authors proposed the use of lasso regression along with the polynomial, which they call as polynomial with lasso.  Lasso regression is a linear model which performs L1 regularization that reduces the number of regression variables and obtains a subset of variables that minimizes the prediction error \cite{Lasso}.  Equation \ref{nineteen} shows the basic function of the polynomial with lasso power model.  In addition, the authors also proposed the use of exponential function along with the polynomial leading to the exponents of linear, quadratic, and cubic functions of the CPU and memory utilization values ($e^{CPU}$,$e^{CPU^2}$,$e^{CPU^3}$,$e^{mem}$,$e^{mem^2}$,$e^{mem^3}$).  Lasso regression is performed in a similar way to reduce the number of estimators.  The basic function of the polynomial + exponential with lasso is stated in Equation \ref{twenty}.  The polynomial with lasso and polynomial + exponential with lasso models were later compared and used by \cite{LUO2013} for energy-efficient scheduling in cloud computing.

\begin{equation}
\label{nineteen}
\phi(.) = \{x_i^a ; 1 \leq a \leq 3\}
\end{equation} 
\begin{equation}
\label{twenty}
\phi(.) = \{e^{x_i^a} ; 1 \leq a \leq 3\}
\end{equation}
where \textit{$x_i$} is the CPU and the memory utilization.

In 2017, \cite{thesis} experimentally found that the power consumption of a server is a function of various S\_PC that corresponds to a server's hardware resources such as the processor, the random access memory, the network interface controller.  The authors used 30 different S\_PC representing the hardware resources to develop the relationship between the server's power consumption and resource utilization.  The exposed low-level system performance counters (S\_PC) are branch-instructions, instructions, cache- misses, L1-icache-load-misses, branch-loads, branch-load-misses, LLC-loads, LLC-store-misses, LLC-load-misses, LLC-stores, dTLB-store-misses, dTLB-load- misses, dTLB-loads, dTLB-stores, bus-cycles, L1-dcache-stores, L1-dcache-load-misses, L1-dcache-loads, CPU cycles, branch-misses, cache-references, iTLB-loads, iTLB-load-misses, node-load, node-stores, node-load-misses, node-stores-misses, ref-cycles, number of if octets out, and number of if octets in.  In order to avoid over-fitting by only selecting the significant counters for a server, the author proposes the use of Lasso regression.  Equation \ref{twenty one} shows the minimization function for the lasso regression model.
\begin{equation}
\label{twenty one}
\mathop{minimize}_{w} \frac{1}{2n_{samples}}||X_{w}-y||^{2}_2 + \alpha||w||_1 \\
\end{equation}

\item \textit{Support Vector Machines (SVM):}
In 2011, \cite{mccullough2011evaluating} found that the CPU and memory utilization values of a server are interdependent and cannot be used in a linear model to accurately predict the power consumption of the server. The authors proposed the use of SVM-based regression model to predict the power consumption using the function stated in Equation \ref{twenty two}.  SVM regression aims at finding a linear hyperplane, that fits the non-linearly correlated multidimensional regression parameters to the output variable \cite{scholkopf2001learning}.  The model is S\_PM-based using CPU and memory utilization as the independent variables.  The model was later evaluated and used by \cite{LUO2013} for energy-efficient scheduling in cloud computing.

\begin{equation}
\label{twenty two}
f(x) = W\phi(x) + b
\end{equation}

where \textit{W} and \textit{b} are the regression parameters calculated using the optimization problem to minimize the function stated in Equation \ref{twenty three} and \textit{$\phi(x)$} is a function of CPU and memory utilization.
\begin{equation}
\label{twenty three}
\mathop{minimize}_{w,b,\zeta} \frac{1}{2}||W||^2 + C\sum_{i=1}^{N}(\zeta_i + \zeta_i^*)
\end{equation}

where C is the error penalty constraint, and $\zeta_i$ and $\zeta_i^*$ are the slack variables bounding the allowable regression errors.

\item \textit{Deep Neural Networks (DNN):}
In 2016, \cite{Li2016} stated that the static power models such as SVLF and SVNLF can not predict the power consumption accurately due to the heterogeneous and dynamic nature of workloads in a data center.  The authors proposed the use of deep neural networks to analyze the trend in the past data center power consumption for prediction.  The proposed deep learning prediction model is based on recursive auto-encoder and uses the power consumption data of a server corresponding to its CPU utilization, CPU load averaged over 5, 10 and 15 minutes, memory utilization, number of disk read/write, packets/s in and out, and  the file system used/available for training the model (S\_PM-based).  The recursive auto-encoder are neural networks that encodes the input into a latent space and tries to reconstruct the input as the output \cite{socher2011semi}.  The auto-encoder output is then used to predict the value of power consumption such that it minimizes the objective function stated in Equation \ref{twenty four}.
\begin{equation}
\label{twenty four}
\epsilon_{RAE} = \epsilon_{PRD}\times 0.95 + \epsilon_{AE}\times0.05
\end{equation}
where \textit{$\epsilon_{PRD}$} and \textit{$\epsilon_{AE}$} are the prediction error and the reconstruction error respectively.  The value of \textit{$\epsilon_{PRD}$} and \textit{$\epsilon_{AE}$} are calculated using Equation \ref{twenty five} and \ref{twenty six}.
\begin{equation}
\label{twenty five}
\epsilon_{PRD} = \frac{\sum_{i=1}^{N_{train}}(||y(t) - y'(t)||)^2}{N_{train}} + 0.0001\times(||W||)^2
\end{equation}
\begin{equation}
\label{twenty six}
\epsilon_{AE} = \frac{\sum_{i=1}^{N_{train}}Err_{REC}(t)}{N_{train}}
\end{equation}
where $N_{train}$ is the size of training data set, \textit{$\epsilon_{PRD}$} is the mean square error of the predicted values using the $L_2$-norm regularization parameter, and $Err_{rec}$ is the reconstruction error.

\item \textit{Artificial Neural Networks (ANN):}
In 2015, \cite{cupertino2015towards} conducted an empirical study showing that the CPU-based linear power models do not provide accurate power prediction, especially for servers  having multicore processor.  This is mainly due to two reasons: 1) the power consumption has a non-linear relationship with the number of cores utilized, and 2) the power consumption of a server is application dependent for a given CPU utilization.  The authors proposed the use of artificial neural networks for the prediction of power consumption.  They used multilayer perceptron (MLP), which is a feedforward ANN composed of one or more hidden layers.  The output of each hidden layer is computed using Equation \ref{temperatureANN}.
\begin{equation}
\label{temperatureANN}
a = \phi(Wi + b)
\end{equation}
where W is the weight matrix, i is in input vector that consists of the independent variables, b is the bias vector, $\phi(.)$ is the activation function and a is the output vector, i.e., the predicted power consumption.

The authors used a MLP model with two hidden layers and a sigmoid activation function.  The power model is based on different S\_PMC variables such as number of instructions, cycles, cache references, cache misses, branch instructions, branch misses, bus cycles, idle cycles frontend, task clock, page faults, context switches, CPU migrations, major and minor faults, L1d loads, L1d load misses, L1d stores, L1d store misses, L1d prefetch misses, L1i load misses, LLC loads, LLC load misses, LLC stores, LLC store misses, L1d prefetches, LLC prefetch misses, dTLB loads, dTLB load misses, dTLB stores, dTLB store misses, iTLB loads, iTLB load misses, branch loads, branch load misses, node loads, node load misses, node stores, node store misses, node prefetches, node prefetch misses, CPU usage, received and sent bytes, and CPU time.  In addition to these variables the model also uses CPU temperature and frequency.

\item \textit{Gaussian Mixture Models (GMM):}
In 2010, \cite{Dhiman} performed experiments to study the power consumption of heterogeneous applications at different CPU utilization levels.  The results show that the relationship between the power consumption and the CPU utilization is not always linear but it is application specific.  The authors found that the power consumption increases linearly with the CPU utilization for an application having high instructions per cycle (IPC), while for an application having high memory access with an increase in the CPU utilization after a certain value, there is no further increase in power consumption.  Moreover, for an application having high cache conflicts, the power consumption decreases after certain CPU utilization value.  Consequently, the authors proposed a power model based on different S\_PMC variables such as CPU utilization, instructions per cycles (IPC), memory access and cache transactions as the independent variables as stated in Equation \ref{twenty seven}.  The prediction is done using Gaussian mixture model (GMM) to dynamically map different clusters of power consumption values with the corresponding clusters of performance metrics.  GMM is a probabilistic model that assumes all the data points of distribution are generated from a mixture of a finite number of Gaussian distributions with unknown parameters \cite{reynolds2015gaussian}.
\begin{equation}
\label{twenty seven}
P = f(CPU,IPC,mem access, cache transactions)
\end{equation}
\end{itemize}

\begin{longtable}{|p{0.1\textwidth}|p{0.11\textwidth}|p{0.71\textwidth}|}
\caption{Limitations of Software-based Power Models.} \label{table:one} \\
\hline
{\bfseries Power Model Equation} & \multicolumn{1}{|c|}{\bfseries Work} & \multicolumn{1}{|c|}{\bfseries Limitations}\\
\hline
\multicolumn{3}{|c|}{\bfseries Mathematical Formula: SVLF [S\_PM - based]}\\ \hline

\ref{one} and \ref{two} & \cite{Heath2006,Fan2007,Qureshi2009,Gmach2009,Buyya2010,Lee2012,Hongyou2013,Raycroft2014,Bagheri2014,Tang2015,Patel2015,Dai2016,Sharma2016,7997707,CHEUNG2018329}& The model is based only on the minimum and the maximum server power consumption values and does not take into consideration the power consumption values of a server's CPU utilization between its idle and full load state.\\ \hline

\ref{three} & \cite{Sinha2011,Beloglazov2012a,Han2016}& The accuracy of the model depends on the ratio of $P_{MIN}$ to $P_{MAX}$.  If the ratio is not close to 0.7, the model gives high value of error.  Moreover, the model does not consider the power consumption values for the CPU utilization values between 0\% and 100\%.\\ \hline

\ref{four}&\cite{Calheiros:2011:CTM:1951445.1951450,Beloglazov2012,Kord2013,Farahnakian2014,Chowdhury2015,Arianyan2015,Rahmanian2017,Wen2017,Kejing2017}& To predict a power consumption value $p$ for a CPU utilization $u$, the model requires the power consumption values $p_1$ and $p_2$ corresponding to the CPU utilization values $u_1$ and $u_2$ respectively, such that $u_1<u<u_2$.\\ \hline

\ref{five} & \cite{Jin2013}& The model requires the power consumption values corresponding to the minimum and the maximum throughput values for each application type.  Moreover, the model does not consider the applications' power consumption behavior between the minimum and the maximum throughput values.\\ \hline

\multicolumn{3}{|c|}{\bfseries Machine Learning: MLLV - SVLR [S\_PM - based]}\\ \hline

\ref{eight} & \cite{Raghavendra2008,Gong2008,Berral2010,Pedram2010,Berral2011,Zhang2013}& The model's accuracy depends on the deviation of the training data set values from the fitted regression line.  Moreover, the model requires calibration for the values of $\alpha$ and $\beta$ for each server architecture type. \\ \hline

\ref{nine} & \cite{Wang2010}&The model's accuracy depends on the deviation of the training data set values from the fitted regression line and on the increment in power consumption for idle server and server with minimum load.  The higher the increment, the more will be the error.  Moreover, the model requires calibration for the values of $\beta$ for each application type on each server architecture type.\\ \hline

\ref{ten} & \cite{Koller2010}& The model's accuracy depends on the deviation of the training data set values from the fitted regression line.  Moreover, the model requires calibration for the values of $\alpha$ and $\beta$ for each server architecture type. \\ \hline

\multicolumn{3}{|c|}{\bfseries Machine Learning: MLLV - MVLR [S\_PM - based]}\\ \hline

\ref{fourteen} &\cite{Economou2006,Smith2010}&  The model's accuracy depends on the deviation of the training data set values from the fitted regression Euclidean hyperplane.  Moreover, the model requires calibration for the values of the regression parameter for each server architecture type.\\ \hline

\ref{fifteen} &  \cite{Alan2014}&The model's accuracy depends on the deviation of the training data set values from the fitted regression Euclidean hyperplane.  The accuracy also depends on the increment in power consumption for idle server and server with minimum load.  The higher the increment, the more will be the error.  Moreover, the model requires calibration for the values of the regression parameter for each server architecture type.\\ \hline

\ref{sixteen} &\cite{Kansal2010,Li2012}& The model's accuracy depends on the deviation of the training data set values from the fitted regression Euclidean hyperplane.  Moreover, the model requires calibration for the values of the regression parameter for each server architecture type.  The model gives a high value of error for network-intensive applications. \\ \hline

\multicolumn{3}{|c|}{\bfseries Machine Learning: MLLV - MVLR [S\_PMC - based]}\\ \hline

\ref{seventeen} &\cite{husain}&  The model's accuracy depends on the deviation of the training data set values from the fitted regression Euclidean hyperplane.  Moreover, the model requires calibration for the values of the regression parameter for each server architecture type. \\ \hline

\ref{eighteen} & \cite{DaCosta2010}& The accuracy of the model depends on how close the data are to the fitted regression model.  The model has high probability of over-fitting the data. Moreover, the model requires calibration for the values of the regression parameter for each server architecture type.\\ \hline

\ref{temperatureMLR2} & \cite{witkowski2013practical}& The accuracy of the model depends on how close the data are to the fitted regression model.  The model has high probability of over-fitting the data. Moreover, the model requires calibration for the values of the regression parameter for each server architecture type.  It requires CPU temperature hardware indicator.\\ \hline

\multicolumn{3}{|c|}{\bfseries Machine Learning: MLLV - MVLR [S\_PC - based]}\\ \hline

\ref{temperatureMLR} &\cite{jarus2014runtime}&  The model's accuracy depends on the deviation of the training data set values from the fitted regression Euclidean hyperplane.  Moreover, the model requires calibration for the values of the regression parameter for each server architecture type and needs to perform different transformations of the performance counters having non-linear relationship with power consumption.  It requires CPU temperature hardware indicator.\\ \hline

\multicolumn{3}{|c|}{\bfseries Mathematical Formula: SVNLF [S\_PM - based]}\\ \hline

\ref{six} &\cite{Fan2007,Qureshi2009,Sun2015PowerEfficientPF}&  The model is based only on the minimum and the maximum server power consumption values without considering the power for a server's CPU utilization values between its idle and full load state. Moreover, the model requires calibration for the value of $r$ for each server architecture type.\\ \hline

\ref{seven} & \cite{Lien2007,Tang2011}&The model is based only on the minimum and the maximum server power consumption values and requires calibration for the values of $\alpha$ and $\beta$ for each server architecture type.\\ \hline

\multicolumn{3}{|c|}{\bfseries Machine Learning: MLNLV - SVPR [S\_PM - based]}\\ \hline

\ref{eleven} &\cite{Janacek2012,Zhang2013}&  The model's accuracy depends on the deviation of the training data set values from the fitted regression polynomial curve.  Moreover, the model requires calibration for the values of the regression parameter for each server architecture type.\\ \hline

 \ref{twelve} &\cite{Guazzone2012}&  The model's accuracy depends on the deviation of the training data set values from the fitted regression polynomial curve.  Moreover, the model requires calibration for the values of the regression parameter for each server architecture type. \\ \hline

\ref{thirteen} & \cite{Zhang2013}& The model's accuracy depends on the deviation of the training data set values from the fitted regression polynomial curve.  The model generally suffers from the issue of over-fitting.  Moreover, the model requires calibration for the values of the regression parameter for each server architecture type. \\ \hline

\multicolumn{1}{|c|}{-} & \cite{canuto2016methodology}& The model's accuracy depends on the deviation of the training data set values from the fitted regression polynomial curve.  The model generally suffers from the issue of over-fitting.  Moreover, the model requires calibration for the values of the regression parameter for each server architecture type. \\ \hline

\multicolumn{3}{|c|}{\bfseries Machine Learning: MLNLV - LR [S\_PM - based]}\\ \hline

\ref{nineteen} & \cite{mccullough2011evaluating,LUO2013}&  The model's accuracy depends on the shrinking of the non-significant variables.  Moreover, the selection of high order polynomial variables may over-fit the model making the prediction less accurate. \\ \hline

\ref{twenty} & \cite{mccullough2011evaluating,LUO2013}&The model's accuracy depends on the shrinking of the non-significant variables.  Moreover, the model has high probability of over-fitting due to the use of exponential along with the polynomial function. \\ \hline

\multicolumn{3}{|c|}{\bfseries Machine Learning: MLNLV - LR [S\_PC - based]}\\ \hline

\ref{twenty one} & \cite{thesis}& The model while shrinking the non-significant variables to zero, does not consider the integrated correlation between those variables and their combined association on the power consumption.\\ \hline

\multicolumn{3}{|c|}{\bfseries Machine Learning: MLNLV - SVM [S\_PM - based]}\\ \hline

\ref{twenty two} &\cite{mccullough2011evaluating,LUO2013}&  The model's accuracy depends on the selection of the kernel and can be computationally complex.\\ \hline

\multicolumn{3}{|c|}{\bfseries Machine Learning: MLNLV - DNN [S\_PM - based]}\\ \hline

\ref{twenty four} &\cite{Li2012}&  The model's accuracy depends on the size of the training data set.  Moreover, the model training process is computationally complex compared to regression based models.\\ \hline

\multicolumn{3}{|c|}{\bfseries Machine Learning: MLNLV - ANN [S\_PMC - based]}\\ \hline

\ref{temperatureANN} &\cite{cupertino2015towards}&  The model's accuracy depends on the size of the training data set, number of hidden layers and the activation function used.  Moreover, the model training process is computationally complex compared to regression based models.  It requires CPU temperature hardware indicator.\\ \hline

\multicolumn{3}{|c|}{\bfseries Machine Learning: MLNLV - GMM [S\_PMC - based]}\\ \hline

\ref{twenty seven} &\cite{Dhiman}&  The accuracy of the model decreases and the computational complexity increases with an increasing number of variables.\\ \hline

\multicolumn{3}{l}{\textsuperscript{*}\footnotesize{\begin{tabular}{l}Single Variable Linear with Fixed Slope and Intercept (SVLF), \\  Machine Learning Linear with Variable Slope and Intercept (MLLV), \\ Single Variable Linear Regression (SVLR), Multi Variable Linear Regression (MVLR), \\ Single Variable Non-Linear with Fixed Slope and Intercept (SVNLF), \\ Machine Learning Non-Linear with Variable Slope and Intercept (MLNLV), \\ Single Variable Polynomial Regression (SVPR), Lasso Regression (LR), Support Vector Machine (SVM),\\ Deep Neural Network (DNN), Artificial Neural Network (ANN), Gaussian Mixture Model (GMM),\\ System\_Performance\_Metrics (S\_PM), System\_performance\_Counters (S\_PC),\\ and System\_Performance\_Metrics\_Counters (S\_PMC)\end{tabular}}}
\end{longtable}

\subsection{Evaluated Works and Software-based Power Model Development Workflow}

\begin{figure}
\centerline{\includegraphics[width=\linewidth]{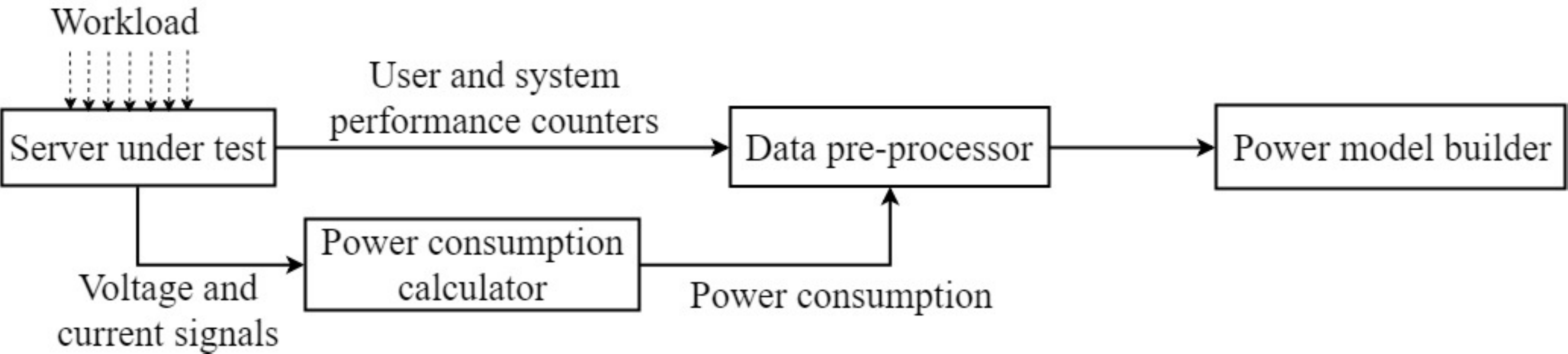}}
\caption{Workflow of power model development.}
\label{fig:one}
\end{figure}

The methodology for power model evaluation comprises of two stages; model development and prediction.  The model development stage generally known as training stage involves building the model based on power consumption values and corresponding performance counters values for some workload or representative benchmark.  The power model development is specific to server architecture requiring a particular model to be trained for each different type of architecture.  Once the model is developed, it is used to predict the power consumption of a server.  This is known as the prediction stage.  Fig. \ref{fig:one} shows the workflow we use to develop the software-based power models under study.  The server under test is the server for which the power models are to be developed.  The workload stressing different user and system performance counters runs on the server.  While the workload is running, the values of the user and low-level system metrics are recorded and written in a file.  Simultaneously, the voltage and the current signal values of the server are measured and sent to the power consumption calculator module.  The values of the counters and the power consumption are then sent to the data pre-processor module, where they are synchronized and averaged to develop the training data set.  This data set is then sent to the power model builder module which builds the power model to be used.  

A similar workflow is used by the works in the literature evaluating different power models.  However, the servers for model development, workload to stress the servers, and the power measurement technique used by these works are different.  For instance, \cite{Pedram2010,Kansal2010,husain,DaCosta2010,Dhiman,Janacek2012,Li2012,Smith2010,LUO2013,Alan2014,thesis,cupertino2015towards} used a single server for power model development, while \cite{Economou2006,Fan2007,Lien2007,Koller2010,Zhang2013,Li2016,CHEUNG2018329,witkowski2013practical,jarus2014runtime,canuto2016methodology} used multiple heterogeneous servers.  Moreover, \cite{Pedram2010,DaCosta2010,LUO2013,thesis} used synthetic workload to stress the server, while \cite{Economou2006,Fan2007,Lien2007,Koller2010,Kansal2010,husain,Dhiman,Smith2010,Li2012,Zhang2013,Alan2014,Li2016,CHEUNG2018329,cupertino2015towards,witkowski2013practical,jarus2014runtime,canuto2016methodology} used different benchmarking applications and real-world workload traces.  In addition, \cite{Lien2007} used DW-6090 power meter, \cite{Pedram2010} used a power analyzer, \cite{Zhang2013} used Chroma 66202 power meter, \cite{Koller2010} used IBM active energy manager, \cite{Janacek2012} used Voltcraft Energy Logger 4000, \cite{Economou2006,Li2012,canuto2016methodology} used a AC power meter, \cite{Alan2014} used Yokogawa WT210 power meter, \cite{Kansal2010,husain} used WattsUp PRO ES power meter, \cite{DaCosta2010} used a watt meter, \cite{LUO2013} used a smart power meter, \cite{jarus2014runtime,witkowski2013practical} used home-brew power meter, and \cite{cupertino2015towards} used a plogg power meter to measure the power consumption of the server.

\graphicspath{{./Images/}}

Table \ref{table:two} summarizes different software-based power models evaluated in the literature.  It shows that these power models are evaluated under different experimental environment, using different error formula, applications and power measurement techniques, which makes it difficult to compare them.  As shown in the table, the errors reported by different works evaluating a similar power model are different.  These discrepancies in the result are due to the use of different experimental environment, setup, and evaluation formula.  To date, there is no work comparing the performance of the models examined in this study.  Thus, in this work, we evaluate these models under a unified experimental environment, power measurement technique and error formula.  We compare the performance on three different server architectures using a diverse set of applications.

\begin{longtable}{ |p{0.06\textwidth}|p{0.07\textwidth}|p{0.12\textwidth}|p{0.11\textwidth}|p{0.1\textwidth}|p{0.1\textwidth}|p{0.08\textwidth}| } 
\caption{Evaluated Works on Software-based Power Models.\label{table:two}}\\
\hline
\multicolumn{1}{|c|}{\bfseries Work}&{\bfseries Power Model Equation}&{\bfseries Experimen- tal Setup} &{\bfseries Workload used to Stress the Server}&{\bfseries Power Measurement Technique}&{\bfseries Error Formula} &\multicolumn{1}{|c|}{\bfseries Error} \\ 
\hline
\hline

\cite{Fan2007}& \ref{one} &Thousand of heterogeneous servers&Webmail, Websearch, and Mapreduce&\multicolumn{1}{|c|}{-}&\multicolumn{1}{|c|}{-}&\multicolumn{1}{|c|}{-} \\ \hline

\cite{CHEUNG2018329}& \ref{one} &Servers from SPECpower 2008 database&SPECpower 2008 database&\multicolumn{1}{|c|}{-}&\multicolumn{1}{|c|}{$\frac{1}{N}\times\sum_{i=1}^{N}\frac{|Predicted-Actual|}{Actual}$}&{25.7\%} \\ 
\hline

\cite{Fan2007}& \ref{six} &Thousand of heterogeneous servers&Webmail, Websearch, and Mapreduce&\multicolumn{1}{|c|}{-}&\multicolumn{1}{|c|}{-}&{1\%} \\ \hline

\cite{Lien2007}& \ref{seven} &Thirteen heterogeneous servers&Streaming media workload using Windows media load simulator&DW-6090 power meter&\multicolumn{1}{|c|}{$\frac{Actual-Predicted}{N}$}&{6\%} \\ \hline

\cite{Pedram2010}& \ref{eight} &Single server&Synthetic workload using workload generator&Power analyzer&\multicolumn{1}{|c|}{-}&{9\%} \\ \hline

\cite{Zhang2013}& \ref{eight} &Seven heterogeneous servers&SPECpower benchmarking application&Chroma 66202 power meter &\multicolumn{1}{|c|}{$\sqrt{\frac{(Actual-Predicted)^2}{N}}$}&{12.95\%} \\ \hline

\cite{Koller2010}& \ref{ten} &Three heterogeneous servers&TPC-W, SPECpower, Domino, daxpy, fma, and HPL applications&IBM active manager&\multicolumn{1}{|c|}{${\frac{|Predicted-Actual|}{Actual}}\times 100\%$}&{5\%} \\ \hline

\cite{Janacek2012}& \ref{eleven} &Single server&\multicolumn{1}{|c|}{-}& Voltcraft Energy Logger 4000& \multicolumn{1}{|c|}{-}&{9\%} \\ \hline

\cite{Zhang2013}& \ref{eleven} &Seven heterogeneous servers&SPECpower benchmarking application&Chroma 66202 power meter&\multicolumn{1}{|c|}{$\sqrt{\frac{(Actual-Predicted)^2}{N}}$}&{7.984\%} \\ \hline

\cite{Zhang2013}& \ref{thirteen} &Seven heterogeneous servers&SPECpower benchmarking application&Chroma 66202 power meter&\multicolumn{1}{|c|}{$\sqrt{\frac{(Actual-Predicted)^2}{N}}$}&{3.319\%} \\ \hline

\cite{canuto2016methodology}& \multicolumn{1}{|c|}{-} &Six heterogeneous servers&Ibench, Stress, Sysbench, Prime 95, Linpack-neon, Pmbw, STREAM, fio, iperf3, CloudSuite, and NAS benchmarks&AC power meter&\multicolumn{1}{|c|}{$\frac{100}{n}\sum_{i=1}^{n}\frac{(Actual-Predicted}{Actual}$}&2.6-5.7\% \\ \hline

\cite{Economou2006}& \ref{fourteen} &Two heterogeneous servers&SPECcpu 2000 integer, SPECcpu2000 floating point, SPECjbb 2000, SPECweb 2005, Streams application, and matrix multiplication&AC power meter&\multicolumn{1}{|c|}{$\frac{1}{N}\times \sum_{i=1}^{N}\frac{Predicted-Actual}{Actual}$}&{4\%} \\ \hline

\cite{Smith2010}& \ref{fourteen} &Four homogeneous servers&Video sharing web application&\multicolumn{1}{|c|}{-}&\multicolumn{1}{|c|}{-}&{3.91\%} \\ \hline

\cite{Alan2014}& \ref{fifteen} &Single server&scp, rsync, ftp, bbcp, and gridftp data transfer tools&yokogawa WT210 power meter &\multicolumn{1}{|c|}{$\frac{1}{N}(\sum_{i=1}^{N}\frac{Predicted-Actual}{Actual})100\%$}&{6\%} \\ \hline

\cite{Kansal2010}& \ref{sixteen} &Single server&SPECcpu 2006, and IOmeter&Power meter WattsUp PRO ES power meter&\multicolumn{1}{|c|}{$\frac{|Predicted-Actual|}{N}$}&{5\%} \\ \hline

\cite{Li2012}& \ref{sixteen} &Six homogeneous servers&pi, sudoku, sort, random writer, and word count Hadoop programs&AC power meter&\multicolumn{1}{|c|}{$\frac{|Predicted-Actual|}{Actual}$}&{4\%} \\ \hline

\cite{husain}& \ref{seventeen} &Single server&NAS-NPB, Iozone, Bonnie++, BYTEMark, Cachebench, Dense matrix multiplication, and Gcc benchmark programs&WattsUp PRO power meter&\multicolumn{1}{|c|}{-}&{94\% accuracy} \\ \hline

\cite{DaCosta2010}& \ref{eighteen} &Single desktop&Synthetic workload&Watt meter&\multicolumn{1}{|c|}{$\sqrt{\frac{(Actual-Predicted)^2}{N}}$}&{2.7\%} \\ \hline

\cite{witkowski2013practical}& \ref{temperatureMLR2} &Three heterogeneous servers&Abinit, NAMD, HMMER, MEncoder and CPU Burn applications, and Intel LINPACK, C-ray and Cavity benchmarks&Home-brew watt meter based around the chipset ADE7763&\multicolumn{1}{|c|}{$\frac{\sqrt{\sum_{i=1}^{N}(Actual - Predicted)^2}}{N}$}&1-4\% \\ \hline

\cite{jarus2014runtime}& \ref{temperatureMLR} &Four heterogeneous HPC servers&Abinit, CPU Burn, HMMER, Namd, MEncoder, FFTE, Make, Mprime, OpenFOAM and Tar applications, and Cavity and C-ray benchmarks&Home-brew device based around the chipset ADE7763&\multicolumn{1}{|c|}{$\frac{1}{N} \sum_{i=1}^{N}(Actual - Predicted)^2$}&1-4\% \\ \hline

\cite{mccullough2011evaluating}& \ref{nineteen} &Single server& * &Data acquisition system and WattsUp meter&\multicolumn{1}{|c|}{$\frac{1}{N}(\sum_{i=1}^{N}\frac{Predicted-Actual}{Actual})$}&{-} \\ \hline

\cite{LUO2013}& \ref{nineteen} &Single server&Synthetic workload&Smart power meter&\multicolumn{1}{|c|}{-}&{10\%} \\ \hline

\cite{mccullough2011evaluating}& \ref{twenty} &Single server& * &Data acquisition system and WattsUp meter&\multicolumn{1}{|c|}{$\frac{1}{N}(\sum_{i=1}^{N}\frac{Predicted-Actual}{Actual})$}&{-} \\ \hline

\cite{LUO2013}& \ref{twenty} &Single server&Synthetic workload&Smart power meter&\multicolumn{1}{|c|}{-}&{10\%} \\ \hline

\cite{thesis}& \ref{twenty one} &Single server&Synthetic workload&\multicolumn{1}{|c|}{-}&\multicolumn{1}{|c|}{$\sqrt{\frac{(Actual-Predicted)^2}{N}}$}&{10\%} \\ \hline

\cite{mccullough2011evaluating}& \ref{twenty two} &Single server& * &Data acquisition system and WattsUp meter&\multicolumn{1}{|c|}{$\frac{1}{N}(\sum_{i=1}^{N}\frac{Predicted-Actual}{Actual})$}&{-} \\ \hline

\cite{LUO2013}& \ref{twenty two} &Single server&Synthetic workload&Smart power meter&\multicolumn{1}{|c|}{-}&{-} \\ \hline

\cite{Li2016}& \ref{twenty four} &Two heterogeneous servers&WC98, and clark web application traces&\multicolumn{1}{|c|}{-}&\multicolumn{1}{|c|}{$\frac{\sqrt{\frac{\sum (Actual-Predicted)^2}{N}}}{Standard deviation}$}&{1.03$\pm$0.13} \\ \hline

\cite{cupertino2015towards}& \ref{temperatureANN} &RECS compute box having eighteen homogeneous computer modules&C0, CU, ALU, FPU, RAND, L1, L2, L3 and RAM micro benchmarks&Plogg power meter&\multicolumn{1}{|c|}{$\frac{100}{N} \times \sum_{i=1}^{N} |\frac{Actual - Predicted}{Actual}|$}&{1.83\%} \\ \hline

\cite{Dhiman}& \ref{twenty seven} &Single server&amm, app, art, cra, eon, equ, fac, fma, gap, gcc, gzi, mcf, mes, per, swi, two, vpr, and wup benchmarks&\multicolumn{1}{|c|}{-}&\multicolumn{1}{|c|}{-}&{10\%} \\ \hline

\multicolumn{7}{l}{\textsuperscript{*}\footnotesize{\begin{tabular}{l}sleep, streamcluster, canneal, memcache, bodytrack, freqmine, x264, blackscholes, stressApp, LinuxBuild, namd, dedup, \\zeusmp, Bonnie, mcf, sphinx3, povray, soplex, cpuload(S\_PMC)\end{tabular}}}

\end{longtable}

\section{PERFORMANCE ANALYSIS}
In this section, we analyze and compare the performance of the studied software-based power models on three different classes of servers, using various tools and benchmarks.  We evaluate their performance using the standard error of estimation. 

\subsection{Experimental Environment}
We use three heterogeneous servers from our Lab located at the College of Information Technology of the United Arab Emirates University, to evaluate the performance of the studied models.  The servers' specifications are listed in Table \ref{table:three}.  We perform various experiments to generate the training, validation and testing data sets.  The power models are developed using the training data set and are then validated using the validation data set.  The models are then evaluated in terms of standard error of estimation using the testing data set.  Table \ref{table:four} shows the list of tools used to generate the training and validation data set.  Table \ref{table:five} shows the list of benchmarks and applications used to generate the testing data set.  We run these tools and applications on each server and measure the values of different resource metrics and corresponding power consumption.  To measure the value of the metrics we use Linux perf utility \cite{Linuxper87:online} and collectd tool \cite{Startpag9:online}, and to measure the corresponding power consumption values, Tektronix – TDS2012B \cite{oscilloscope} 100 MHz with 1GS/s of sampling (2-channel digital oscilloscope) was used.  We connect the oscilloscope to a current probe \cite{wedlock1969electronic} and a high differential voltage probe \cite{wedlock1969electronic} to measure the current and the voltage signals respectively.  The power consumption is then the product of the measured current and voltage signals.  We also use the servers from SPEC power \cite{SPECpowe70:online} to evaluate the performance of the power models in order to verify our evaluation on modern server architectures.  We only evaluate single variable power models considering CPU utilization as the independent variable for the SPEC power servers because only the data for power consumption corresponding to CPU utilization is available on the SPEC power website.  We use two servers listed in the SPEC power results for quarter 1 of 2019, whose specifications are listed in Table \ref{table:six}.  The servers from the SPEC Power website belong to the same family of servers as the ones we use in our experimental testbed, but with different architectures and capabilities.

\begin{table}
\caption{List of Servers Used in the Experiments.}
\label{table:three}
\begin{tabular}{|c|p{0.8\textwidth}|} 
\hline
Server 1 & CELSIUS R940power 2 x Intel Xeon E5-2680v4 CPU (2.40 GHz, 14 cores), 8 x 32GB DDR4, 2 x HDD SAS 600GB, OS version Redhat Enterprise Linux Server RHEL 7.4 – 64-bit. \\
\hline
Server 2 & Sun Fire Intel\_Xeon CPU core of 2.80 GHz, Dual core, with 512 KB of cache and 4 GB of memory for each core, OS version CentOS 6.8(i686). \\ 
\hline
Server 3 & Sun Fire X4100 with AMD\_Operaton252 CPU of 2.59 GHz, dual CPU, single core, with 1MB of cache and 2GB of memory for each core, OS version Red Hat Enterprise Linux Server release 7.3 (Mapio). \\ 
\hline
\end{tabular}
\end{table}
\begin{table}
\caption{Tools Used to Generate the Training and Validation Data sets.}
\label{table:four}
\begin{tabular}{ |p{0.1\textwidth}| p{0.1\textwidth} | p{0.7\textwidth}|}
\hline
{\bfseries Tool} & {\bfseries Resource Stressed} & {\bfseries Description} \\ 
\hline
\hline
CPU Load Generator \cite{GitHubGa14:online}& CPU & CPU Load Generator is a script written in Python that allows generating a fixed CPU load for a finite user defined time duration.  The script takes in as input the desired CPU load, the duration of the experiment and the CPU core on which the load must be generated. \\\hline 
{Stress \cite{stresspr3:online}} & Memory & Stress is used to generate a configurable measure of CPU, memory and disk load.  We use Stress-1.0.4 to generate configurable stress on memory.  The inputs to the stress command line are the number of vm workers, memory allocation size per vm worker and the duration of the experiment. \\\hline 
{Vdbench \cite{Vandenbergh2012}} &Disk I/O rate& Vdbench is used to generate configurable amount of disk I/O workloads on a system.  We set the desired I/O rate using the curve parameter of the run definition file. \\\hline
{iperf3 \cite{iperfa,iPerfDow4:online}} &Network I/O rate& Iperf3 is a tool for active measurement of the maximum available bandwidth on IP networks.  We use iperf3 between to generate a configurable network I/O rate between the test server and a remote host server. \\
\hline
\end{tabular}
\end{table}
\begin{table}
\caption{Applications and Benchmarks Used to Generate the Testing Data set.}
\label{table:five}
\begin{tabular}{ | p{0.28\textwidth} | p{0.1\textwidth} | p{0.52\textwidth}|} 
\hline
{\bfseries Application/Benchmark} & {\bfseries Resource Stressed} & {\bfseries Description} \\ 
\hline
\hline
Sysbench benchmark \cite{Kopytov2006,GitHubak75:online}& CPU & Sysbench benchmark is used to evaluate the OS parameters like CPU utilization, memory utilization, and Disk I/O.  We use the Sysbench to stress the CPU, using the CPU workload \cite{Sysbench92:online}.  The CPU workload calculates prime numbers between zero and a specified number. \\ \hline 
MEncoder application \cite{MPlayerT28:online} & CPU and Memory & We used MEncoder 4.45, a video compressor application included in the Mplayer project, to stress the CPU and the memory.  We use the MPEG-4 video format \cite{WhatisH287:online}, with 1920 x 1080 resolution, 18,356.7 kbps, 23 fps, and 24 bpp. \\ \hline 
PARSEC benchmark -Black Scholes Model (Portfolio management) \cite{Bienia2011,ThePARSE52:online}& CPU, Memory and Disk &The Black Scholes by Intel RMS benchmark calculates the prices of European options' portfolio analytically using the partial differential equation (PDE). \\ \hline 
Data Mining - Ensemble Clustering application \cite{Weka3Dat77:online} & CPU, Memory and Disk & We use Weka 3.8 \cite{Weka3Dat77:online} to perform k-means clustering of the forest cover \cite{UCIMachi73:online} data set consisting of Geo-spatial descriptions of various forest's types.  The data contains 581,000 instances, 7 classes, and 54 attributes. \\\hline
PARSEC benchmark -Streamcluster  \cite{ThePARSE52:online,Bienia2011}& CPU, Memory, Disk and Network & Streamcluster is a part of the PARSEC 3.0 benchmark suite to solve the online clustering problem.  Stream clustering is memory intensive for low-dimensional data and becomes CPU intensive as the dimension increases.\\
\hline
\end{tabular}
\end{table}

\begin{table}
\caption{List of Servers from SPEC Power Used in the Experiments.}
\label{table:six}
\begin{tabular}{|c|p{0.8\textwidth}|} 
\hline
SPEC\_Server 1 & Dell PowerEdge R7425, AMD EPYC 7601 2.20 GHz, 32 core, 64 MB L3 cache, 16x8 GB of memory, 240 GB SATA SSD, and OS Microsoft Windows Server \cite{SPECpowe33:online}. \\
\hline
SPEC\_Server 2 & Lenovo Global Technology ThinkSystem SR150, Intel Xeon E-2176G, 6 core, 3.7 GHz, 12MB L3 cache, 2x16 GB of memory, 128 GB M.2 SSD, and OS Microsoft Windows Server \cite{SPECpowe10:online}. \\ 
\hline
\end{tabular}
\end{table}

\subsection{Experiments}
The set of experiments performed on the servers to obtain the training, validation, and testing data sets for the power model development and performance evaluation are discussed in this section.  The performance of the different user and system counters used by the studied power models was measured in real-time using the Linux tools.  We also measure the corresponding power consumption values using a LabVIEW program.  The values of the counters and power consumption are written to a file every one second and are then averaged.  We repeat all the experiments 5 times and averaged the averages.

For all the models under study, except for throughput-based S\_PM model, the experiments for generating the training and validation data sets are performed by stressing the CPU, memory, disk operations, and network transfers individually on each of the three servers.  We stress the CPU by generating a CPU load between 0\% - 100\% for 5 minutes each at random intervals, using the CPU Load Generator tool.  For multi-core servers, we generate the CPU load on all the cores simultaneously.  We use the Stress tool to populate the memory using random memory sizes for a virtual machine (vm) worker, for 5 minutes each.  To stress the disk I/O at configurable I/O rate, we then use vdbench tool.  We first find the maximum disk I/O rate of the server and then generate I/O rates between 0\% - 100\% of the maximum I/O rate for 5 minutes each.  We stress the network I/O rate by specifying the desired network bandwidth between the test server and a remote desktop.  We measure the maximum available bandwidth for the server using iperf3 and then ping the remote desktop with random bandwidth between 0\% - 100\% of the maximum bandwidth for 5 minutes each.

For the throughput-based S\_PM power model, we generate the training and validation data sets as follows.  We use different tools to mimic different resource-intensive applications.  We measure the maximum throughput that can be achieved by an application and run it with random throughput varying between the minimum and maximum.  We use the CPU load of Sysbench benchmark to mimic a CPU- intensive application with throughput represented as the floating operations per second.  For disk-intensive applications, we use vdbench tool with the throughput represented as the number of disk reads/writes per second.  For the network-intensive applications, we use iperf3 tool with the throughput represented as the number of data transferred/received per second.  For the evaluation of the power models based only on CPU utilization for the servers from SPEC power, we used the SPEC power results of power consumption corresponding to different level to CPU utilization for each of the SPEC power servers.

For all the models under study, the training and validating data sets set are selected randomly using 70\% and 30\% of the generated experimental data set respectively.

To generate the testing data set, we run Sysbench for the CPU workload to calculate the prime numbers up to 20000000 with the number of threads increasing randomly from 0 up to the total number of threads of the server under test.  In addition, we run the MEncoder application to compress a video file of size 100MB.  We repeat the process for video files of sizes 200MB-2GB, with an increment of 100MB.  Furthermore, we use the Black Scholes application to calculate the prices of a 65,536 European options portfolio.  We also use the ensemble clustering application to perform k-means clustering of data sets with a different number of instances.  We use 4 different sizes of 7.38MB, 74.2MB, 746MB, and 941MB having 27900, 279000, 2790000, and 5580000 instances respectively, form the UCI Forest data repository.  Moreover, we run the Streamcluster application from the PARSEC benchmark suite to perform online stream clustering for native input options having 1,000,000 input points and 128 dimensions.  The power models performance for the servers from SPEC power is not evaluated as the testing data set using different benchmarking applications for those servers can not be obtained as they are not part of our experimental testbed.

We use the R programming language \cite{RTheRPro66:online} to develop the studied power models using the generated training data set and to evaluate their performance using the validation and testing data sets.  The performance of the models is analyzed using standard error of estimation calculated using Equation \ref{twenty eight}.
\begin{equation}
\label{twenty eight}
e_{est} = \sqrt{\frac{\sum_{i=1}^{n} (P_i - P_i' )^2}{n}}
\end{equation}
Where \textit{P} and \textit{P'} are the actual and predicted values of power consumption respectively and \textit{n} is the length of the testing data set.

\subsection{Experimental Results Analysis}
In this section, we discuss the results obtained by the works on software-based power models when evaluated by those works under the same experimental environment and setup and compare them with our results.  We also analyze our experimental results and give insights and conclusions of these evaluations.  In particular, we reveal the reasons behind the performance of these models.

\subsubsection{Analysis of the Evaluated Works on Software-based Power Models in the Literature}

\paragraph{}
Table \ref{table:pastresult} shows the results for prediction errors of different power models obtained by the works in the literature using a unified setup.  \cite{Zhang2013} evaluated the SVLR model in Equation \ref{eight} and SVPR models in Equation \ref{eleven} and Equation \ref{thirteen} and reported an error of 12.95\%, 7.98\% and 3.32\% respectively.  These results indicate that 3rd order SVPR model (Equation \ref{thirteen}) outperforms the 2nd order SVPR (Equation \ref{eleven}) and SVLR (Equation \ref{eight}) models.  \cite{LUO2013} evaluated and compared the Equations \ref{nineteen}, \ref{twenty}, and \ref{twenty two} and showed that the SVM model in Equation \ref{twenty two} has the least error, while the model in Equations \ref{nineteen} and \ref{twenty} has almost the same error.  The GMM model in Equation \ref{twenty seven} was evaluated and compared to the SVLR and MVLR models by \cite{Dhiman}.  The results showed that GMM has an error of 10\%, compared to SVLR model in Equation \ref{eight} having an error of 50\%.  To our knowledge, there is no work which compares the software-based power models in the literature.  In the sections that follow, we compare and analyze these models using a unified experimental setup, workload, and error calculation formula.

\begin{table}
\caption{Results of Power Prediction Errors Obtained by the Works Comparing Models in the Literature}
\label{table:pastresult}
\begin{tabular}{|c|c|c|} 
\hline
{\bfseries Work} & {\bfseries Power Model Equation} & {\bfseries Error} \\ 
\hline
\hline

\multirow{3}{*}{\cite{Zhang2013}} & \ref{eight} & 12.9\% \\ \cline{2-3}
& \ref{eleven} & 7.9\% \\ \cline{2-3}
& \ref{thirteen} & 3.3\% \\ \hline

\multirow{3}{*}{\cite{LUO2013}} & \ref{nineteen} & 10.0\% \\ \cline{2-3}
& \ref{twenty} & 10.0\% \\ \cline{2-3}
& \ref{twenty two} & <10.0\% \\ \hline

\multirow{2}{*}{\cite{Dhiman}} & \ref{eight} & 50.0\% \\ \cline{2-3}
& \ref{twenty seven} & 10.0\% \\ \hline

\end{tabular}
\end{table}

\subsubsection{Analysis of our Experimental Results}

\paragraph{}
In this section, we evaluate the performance of the power models under study for the generated validating data sets and the testing data set.  In particular, we compare the performance among the CPU-based models, among the throughput-based models, and among the multi variable models for the validating data sets.  In addition, we compare the performance among all the models for the testing data set.
 
\begin{figure}
\centerline{\includegraphics[width=\linewidth]{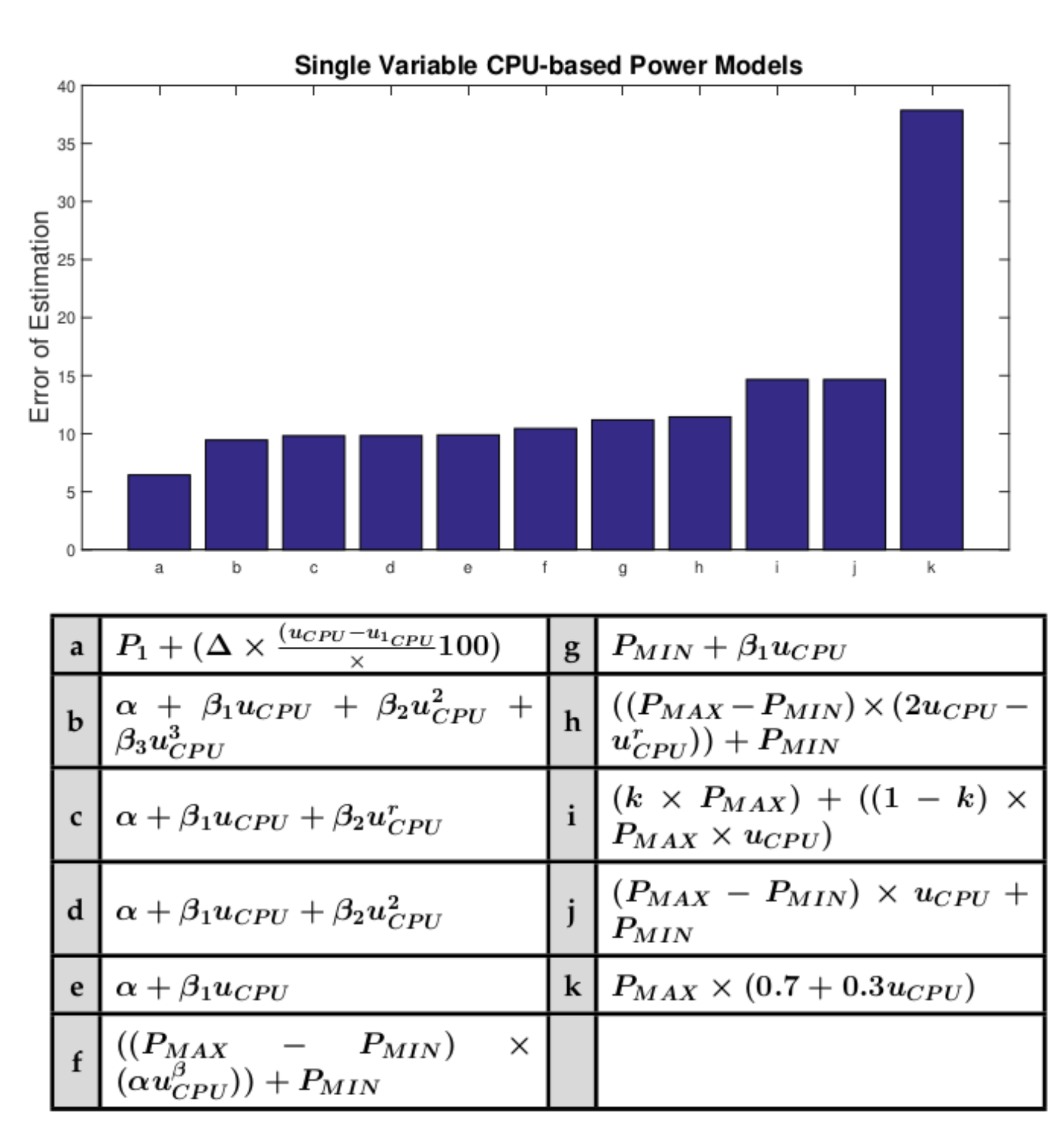}}
\caption{Error of estimation of single variable CPU-based power models for validation data set.}
\label{fig:CPU}
\end{figure}

\subsubsubsection{Single Variable CPU-based Power Models for Validating Data Set}
Figure \ref{fig:CPU} shows the standard error of estimation of the single variable CPU-based power models for the validating data set for Server 1.  It shows that the interpolation model has the least error of estimation, followed by the models based on SVPR, SVLR, SVNLF, and SVLF.  This is thanks to the piece-wise linearization between every two data points of the training data set resulting in a better prediction.

Comparing the performance of SVPR with SVLR models, the error of estimation with the SVPR models is less, which is also confirmed by the evaluation results in the literature (Table \ref{table:pastresult}).  This is because, for a server, the power consumption profile corresponding to the CPU utilization values fits well to a curve rather than a linear line.  Among the SVPR models, the 3rd order has the least error compared to the 2nd and rth order.  This is because the power consumption behavior of the server is an increasing function at the endpoints which can be more accurately represented using a 3rd-degree polynomial curve with the end-points moving in the same direction.  Whereas, in a 2nd-degree polynomial curve, the endpoints move in the opposite direction resulting in a higher error.

The error of SVLF and SVNLF power models is more compared to that of SVLR models (Figure \ref{fig:CPU}).  This is because the SVLF and SVNLF are based only on the endpoint power consumption values, $P_{MAX}$ and $P_{MIN}$, to construct a line where all the possible predicted values will lie.  Therefore, they do not consider the implications of other power consumption data between the endpoints for predictions.  However, the SVLR models compute a linear regression line to best fit the data distribution while minimizing the sum of the squares of the vertical regression deviations.  For the SVLR, the model with the fixed intercept ($P_{MIN}$) has a higher error compared to the model with dynamic intercept.  This is because there is a sudden change of slope in the power consumption trend for the CPU utilization value at 0\% and at values greater than 1\%.  This change is not captured by the regression model having fixed intercept, consequently having more prediction error.

Comparing the performance of SVLF with SVNLF power models, the SVLF has more error.  This is because SVLF models do not capture slight non-linearity of the power data distribution over the range of CPU utilization values.  Consequently, the straight line computed by the SVLF models for predictions has a high value of offset compared to the SVNLF models.  The SVLF model assuming that $P_{MIN}$ is 70\% of $P_{MAX}$ has the maximum error among all the single variable power models.  The rationale behind that is this assumption, which does not hold true for each class of the server.  The higher the deviation from the assumption, the higher will be the error.

\begin{figure}
\centerline{\includegraphics[width=\linewidth]{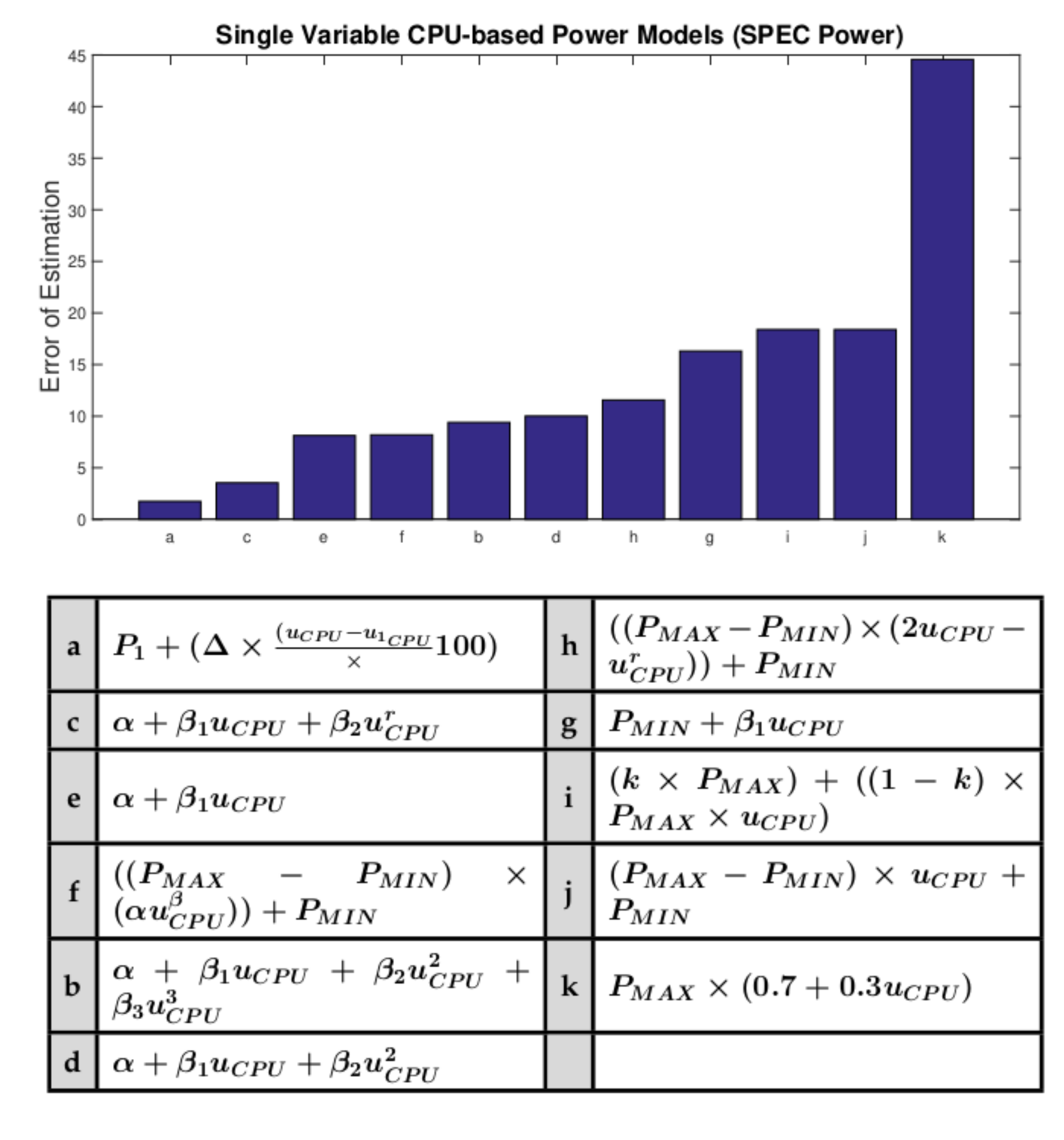}}
\caption{Error of estimation of single variable CPU-based power models for SPEC power data set.}
\label{fig:spec}
\end{figure}

Figure \ref{fig:spec} shows the error of estimation of the single variable CPU-based power models for the SPEC power servers.  The performance of the models remains the same as that of the servers used in our experimental testbed.  However, the performance of the SVLR is better than that of SVPR models using the SPEC power servers.  This is because the power consumption profile is linear with the CPU utilization for SPEC power servers while for our servers the power profile fits better to polynomial curve than a linear line.  This indicates that the performance relative performances of SVLR and SVPR depend on the server power profile.  The performance of the remaining models which are CPU-based is the same on SPEC Power and our experimental testbed (Figures \ref{fig:CPU} and \ref{fig:spec}).

\begin{figure}
\centerline{\includegraphics[width=\linewidth]{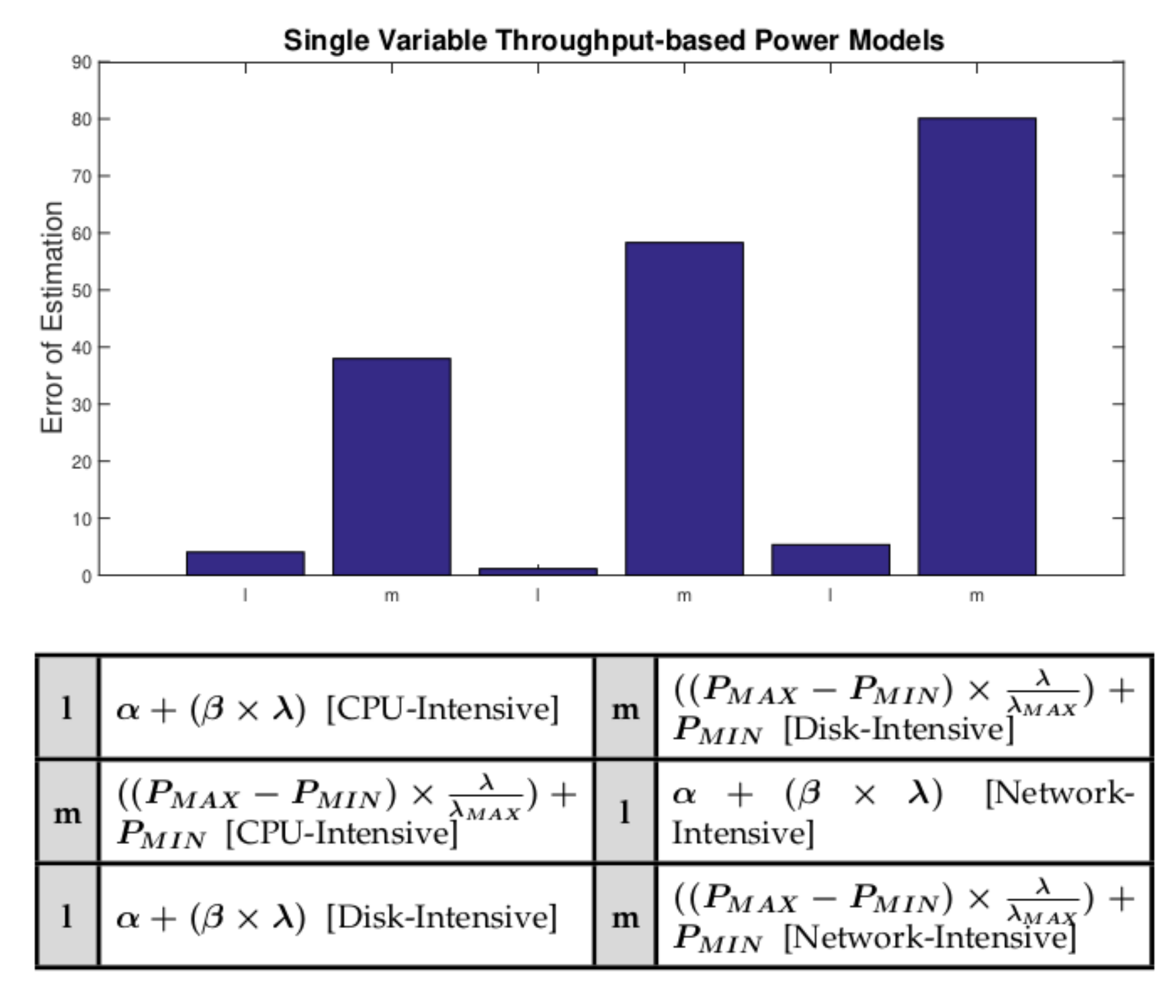}}
\caption{Error of estimation of single variable throughput-based power models for validation data set.}
\label{fig:throughput}
\end{figure}

\subsubsubsection{Single Variable Throughput-based Power Models for Validating Data Set}
Figure \ref{fig:throughput} shows the error of estimation for the throughput-based power models when evaluated using the validation data set generated by running CPU, memory and disk-intensive applications with varying throughput.  It shows that for all application types the error of estimation of SVLF is greater than that of the SVLR model.  This is because SVLF models only consider the power consumption values of the endpoints CPU utilization and do not take into consideration the spatial distribution and behavior of the power data.  The SVLR model based on throughput also outperforms the models based only on CPU utilization, except the interpolation model.  This is because the throughput based model considers the impact of each underlying resource of the server environment that contributes to the power consumption.

\begin{figure}
\centerline{\includegraphics[width=\linewidth]{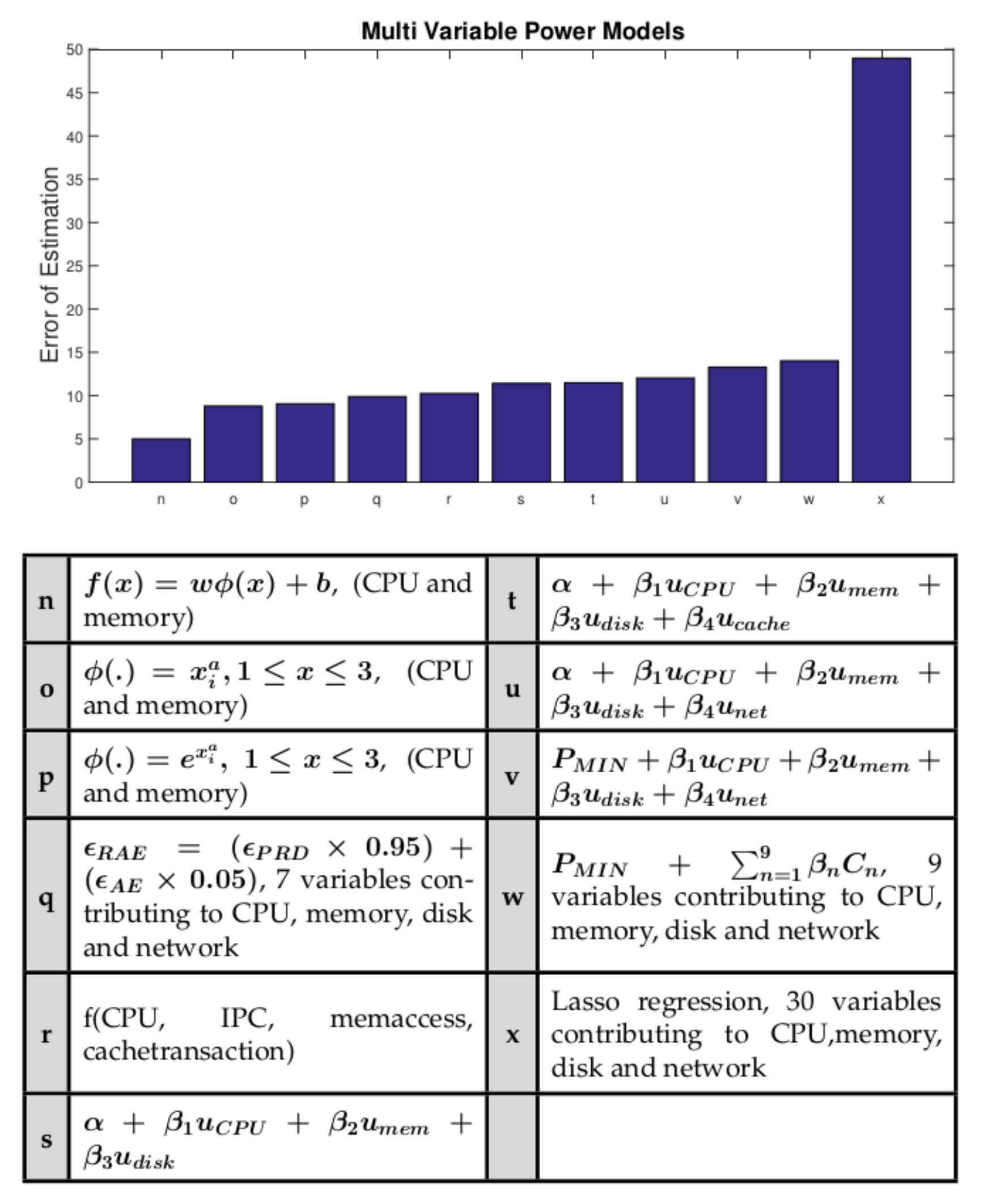}}
\caption{Error of estimation of multi Variable power models for validation data set.}
\label{fig:Multi}
\end{figure}

\subsubsubsection{Multi Variable Power Models for Validating Data Set}
Our results (Figure \ref{fig:Multi}) for the standard error of estimation of multi variable power models for the validating data set for Server 1 shows that the SVM model based on CPU and memory utilization has the least error of estimation compared to other evaluated multi variable power models.  This is because of the server's non-linear power profile and cross-dependency between the variables.  SVM considers the variable dependency and transforms the non-linear data into a high dimensional feature space acknowledging the presence of non-linearity and gives precise predictions compared to other multi variable linear models where the variables are assumed to be independent.  The non-linear polynomial and polynomial+exponential power models with lasso have almost similar performance with the second least error after SVM.  These models consider the quadratic and cubic functions of the CPU and memory utilization values resulting in a regression hyperplane that fits close to the actual values.  Consequently, these models have less error compared to the models considering only the linear functions of resource utilization.

The error of estimation of DNN model is less compared to that of the power models based on GMM and MVLR models.  This is thanks to the use of recursive autoencoder in the neural network power model.  The recursive autoencoder model generates an encoder output as a function of the current data point and previous encoder output.  Consequently, the recursive autoencoder will generate a dynamically varying prediction line as a time series minimizing the prediction offset.  The better performance of GMM compared to the MVLR models (also confirmed by the evaluation in the literature as stated in Table \ref{table:pastresult}) is because that GMM considers the interaction of different variables resulting in various levels of power consumption instead of having a single linear hyperplane representing the power.

Comparing the performance of different MVLR models, the power models based on CPU, memory and disk has less error compared to the models including network and cache in addition (Figure \ref{fig:Multi}).  This is because the inclusion of variables that are not significant for power consumption will over-fit the regression model causing a high offset between the fitted and the actual values.  The model including the cache in addition to CPU, disk and memory has less error compared to the network inclusive model.  This is because the cache transaction is reflected by the utilization of memory contributing to the power consumption and thus yielding accurate predictions than the model including network instead of cache.  The error of MVLR model with fixed intercept is more compared to the MVLR models with dynamic intercept having at most 4 independent variables.  The rationale behind this is the sudden change of slope in the power consumption trend for an idle and utilized server, which is not modeled when using a fixed intercept.  The performance of the throughput-based models (Figure \ref{fig:throughput}) is better than the MVLR models because the MVLR models do not capture all the underlying performance counters that contribute to the power which is however captured by the application throughput. 

Figure \ref{fig:Multi} shows that the MVLR model with 9 independent variables has the second-highest error and the LR model with 30 variables has the highest error among all the evaluated models.  This is because the model with 9 variables includes context switch and interrupt requests, which do not contribute to the power consumption majorly.  The worst performance of the LR model with 30 variables is because instead of considering the CPU utilization, the model takes 28 different low-level performance counters contributing to CPU utilization, each of them contributes to the power consumption.  The lasso regression only selects some significant power contributors while shrinking the remaining.  Consequently, the model leads to a high prediction error as the relationship between the rejected metrics and the power consumption is not modeled.

In summary, interpolation model has the least error of estimation for the single variable power models when evaluated using the validation data set, while the model assuming the idle power to be 70\% of the server's peak power has the maximum error of estimation.  For the multi variable power models, SVM has the best performance with least error, while the lasso regression model with 30 variables has the maximum error of estimation.  The errors of estimation for interpolation, a model assuming idle power as 70\% of peak power, SVM and lasso regression with 30 variables are 6.44, 37.85, 4.98 and 48.98 respectively.  Our experiments show that the relative performance of the models remains the same for servers 2 and 3 used in the experimental setup.

\begin{figure}
\centerline{\includegraphics[width=\linewidth]{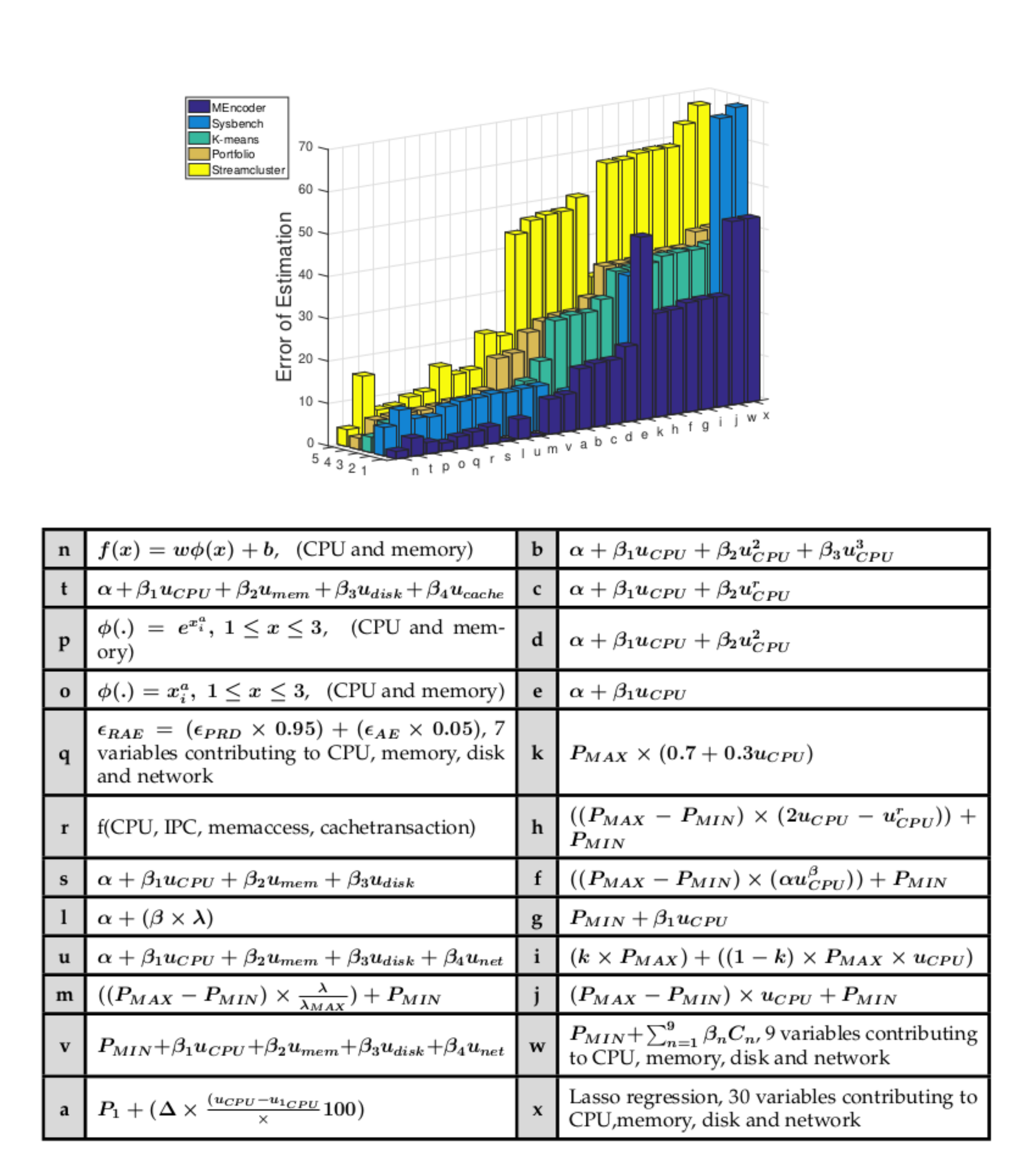}}
\caption{Error of estimation of software-based power models using testing data set for different applications.}
\label{fig:Applications}
\end{figure}

\subsubsubsection{Software-based Power Models for Testing Data Set}
Figure \ref{fig:Applications} shows the performance of all the software-based power models when evaluated for the testing data set, i.e., real applications.  For the CPU-intensive Sysbench application, it is expected that the single variable power models considering CPU utilization should perform better than the multi variable models.  Our results (Figure \ref{fig:Applications}) show that the interpolation model has the least error of estimation compared to the other models because of its piece-wise linearization approach as discussed previously.  The SVM, polynomial with lasso and polynomial+exponential with lasso models still outperforms other models in terms of error of estimation.  This is because of the acknowledgment of non-linearity by the SVM model and the inclusion of quadratic and cubic functions by the polynomial and polynomial+exponential models.  Comparing the performance of MVLR models having at most 4 independent variables with SVLF and SVNLF models, the non-regression models considering only the power consumption values at the end points have more error of estimation.  The LR model with 30 variables has the worst performance with the maximum error of estimation compared to other evaluated models.

Regarding the CPU and memory-intensive application MEncoder, the performance of multi variable models is better than that of the single variable models.  This is thanks to the inclusion of memory utilization while modeling the power consumption by the multi variable models.  For the MEncoder application, SVM model has the least error while the lasso regression model with 30 variables has the maximum error of estimation.  Figure \ref{fig:Applications} shows that for CPU, memory and disk-intensive K-means application, the overall performance of all the models remains the same except the relative performance of MVLR model with CPU, memory and disk, and model with CPU, memory, disk and cache.  It shows that the model including the cache has less error because K-means application generates cache references contributing to power consumption considered by the power model, thus leading to more accurate predicted values.  For the portfolio application with more cache transactions, the multi variable regression model with cache outperforms the polynomial with lasso and polynomial+exponential with lasso models, with SVM having the least error while the lasso regression model with 30 variables having the maximum error.

\begin{figure}
\centerline{\includegraphics[width=90mm]{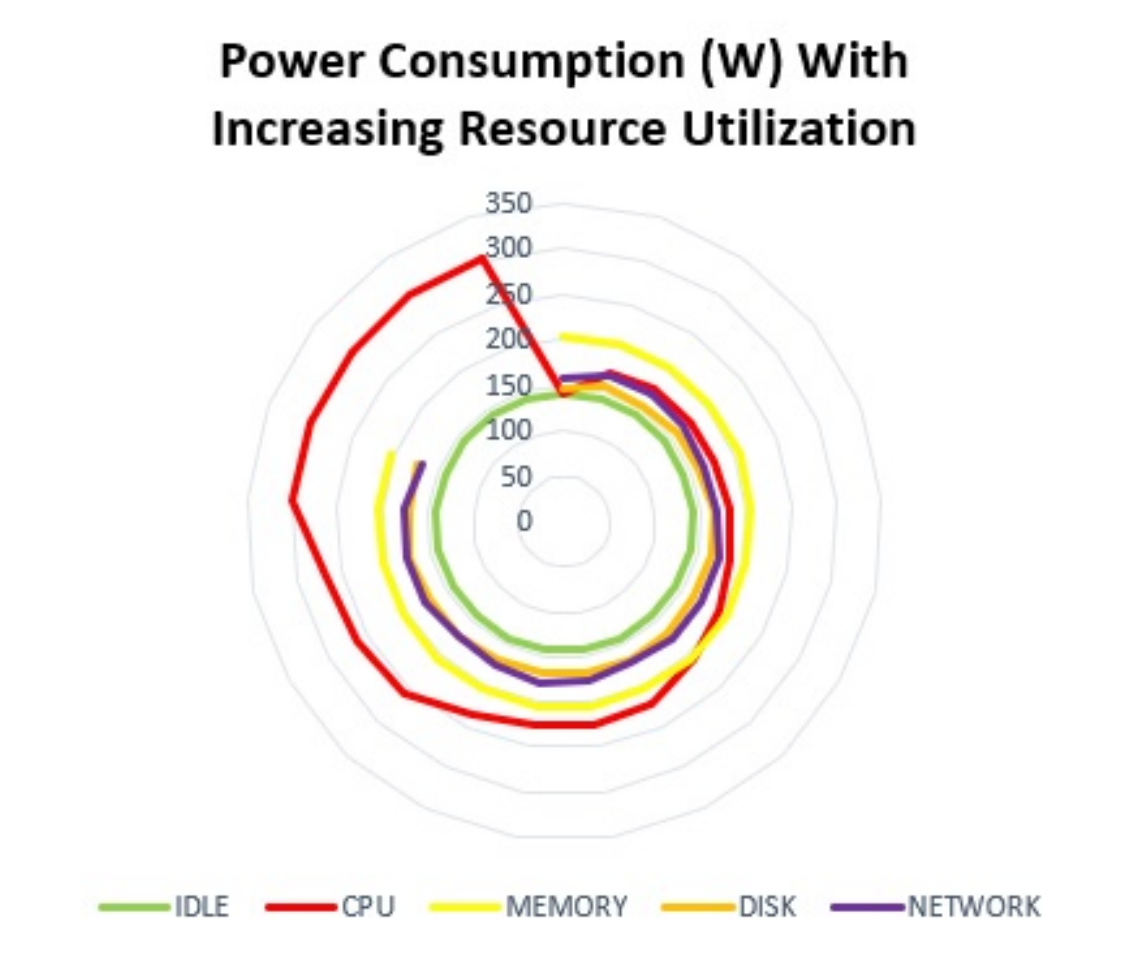}}
\caption{Power consumption (W) of server 1 for increasing utilization of CPU, memory, disk and network.}
\label{fig:POWER}
\end{figure}

Comparing the performance of the MVLR models with at most 4 independent variables, the model that includes network utilization has the least error of estimation for the Streamcluster application.  This is because the application performs online clustering utilizing the network contributing to a small amount of power consumption (Figure \ref{fig:POWER}).  Consequently, the model with network utilization in addition to CPU, memory and disk models the relationship between resource utilization and power consumption more precisely, leading to less error.  The overall performance of other models still remains the same for the Streamcluster application.

The error of estimation for the power models based on CPU utilization is higher than the multi variable models and application's throughput-based models.  This is because the correlation between the power and the performance counters other than CPU is not considered in the models based only on CPU utilization.  Figure \ref{fig:POWER} shows the power consumed by server 1 with increasing values of CPU, memory, disk and network utilization.  It shows that the power consumption of the server is dominated by CPU utilization.  It shows that at 100\% CPU load the server consumes 302W of power.  However, the memory consumes 200W independent of the memory load.  This consumption is higher the server power consumption at idle state.  The increase in power consumption with disk and network utilization is not significant.  The maximum power consumption for 100\% disk utilization is 173W and with 100\% network utilization is 178W.  Consequently, the models not considering memory utilization have a high error of estimation compared to the ones considering the memory.

The results obtained in our experiments can not be compared with that obtained in the past due to discrepancies in the experimental setup, environment, and workloads.  However, we compare the relative performance of the models that are evaluated under the same setup in the past with the results from our experiments.  Similar, to evaluation result by \cite{Zhang2013} the relative performance of the models in Equations \ref{eight}, \ref{eleven}, and\ref{thirteen} remains the same in our results.  Equation \ref{thirteen} has the least error of estimation compared to Equations \ref{eight} and \ref{eleven}.  The relative performance of Equations \ref{nineteen}, \ref{twenty} and \ref{twenty two} reported by \cite{LUO2013} is also confirmed in our experimental results with SVM model (Equation \ref{twenty two}) having the least error of estimation.  According to the results obtained by \cite{Dhiman}, GMM model (Equation \ref{twenty seven}) has least error compared to the SVLR and MVLR models.  This is also confirmed in our results.  The results of the works evaluating single power models in the past can not be compared with the results in the literature and with our results.  This is due to the use of different experimental setup, environment, power measuring technique, error calculation formula, and workloads.  Thus in this paper, we evaluated the software-based power models in a unified setup to have a qualitative comparison between them.

In summary for the Sysbench application, interpolation has the least error of 2.60, while the lasso regression model with 30 variables has the maximum error of 68.98.  For the MEncoder, K-means, portfolio and Streamcluster applications, the SVM has the least error of 1.78, 3.66, 2.70, and 4.02 respectively, while the lasso regression model with 30 variables has the maximum error of 43.48, 36.60, 39.11, and 67.23 respectively.  Our experiments reveal that the relative performance of the models remains the same for servers 2 and 3.

\section{RELATED WORKS}
In the last decade, there have been many research efforts both by the academic and the industrial researchers aimed at reducing the computing infrastructure's energy consumption from the circuit level to the data center level.  Power consumption modeling at different levels in a data center has then been proposed in the past for energy efficient designing and optimization, to curb the increasing energy consumption. Several works have proposed power models to be used either for simulation as a tool in designing energy-efficient data centers \cite{Li2003,Heath2006,Economou2006,Fan2007,Raghavendra2008,Pedram2010,Nagasaka2010,Davis2012,Janacek2012}, or for server-level optimization \cite{Li2003,Economou2006,Fan2007,Qureshi2009,Buyya2010,Pedram2010,Nagasaka2010,Liu2011,Lee2012,Beloglazov2012a,Davis2012,Hongyou2013,Bagheri2014,Raycroft2014,Alan2014,Dai2016,Sharma2016,7997707}.  The power models can be classified as: hardware-based, using variables such as server fan speed, voltage, current, capacitor, motherboard components, and resistance for modeling \cite{elnozahy,Wu,Shin2009,7849452,Sarood2014,Mills,Ham2015} and software-based, using variables such as utilization of CPU, memory, disk, network, throughput, interrupts, cache transactions and disk file system for modeling \cite{Heath2006,Economou2006,Fan2007,Lien2007,Raghavendra2008,Gong2008,Gmach2009,Qureshi2009,Buyya2010,Berral2010,Pedram2010,Wang2010,Koller2010,Kansal2010,husain,DaCosta2010,Dhiman,Sinha2011,Calheiros:2011:CTM:1951445.1951450,Tang2011,Berral2011,Lee2012,Beloglazov2012a,Beloglazov2012,Li2012,Janacek2012,Guazzone2012,Smith2010,Zhang2013,Kord2013,Hongyou2013,Jin2013,Bagheri2014,Raycroft2014,Farahnakian2014,Alan2014,Tang2015,Arianyan2015,Chowdhury2015,Sun2015PowerEfficientPF,Dai2016,Sharma2016,Han2016,Li2016,7997707,Rahmanian2017,Wen2017,Kejing2017,thesis,CHEUNG2018329,bircher2011complete,canuto2016methodology,jarus2014runtime,witkowski2013practical,cupertino2015towards}.  In this paper, we focused on software-based computing server's power models.  These models include modeling based on different linear \cite{Heath2006,Fan2007,Qureshi2009,Gmach2009,Buyya2010,Sinha2011,Calheiros:2011:CTM:1951445.1951450,Lee2012,Beloglazov2012a,Beloglazov2012,Hongyou2013,Jin2013,Kord2013,Farahnakian2014,Bagheri2014,Raycroft2014,Tang2015,Chowdhury2015,Arianyan2015,Dai2016,Sharma2016,Han2016,7997707,Rahmanian2017,Wen2017,Kejing2017,CHEUNG2018329,Raghavendra2008,Gong2008,Berral2010,Pedram2010,Wang2010,Koller2010,Berral2011,Zhang2013,Economou2006,Kansal2010,husain,DaCosta2010,Li2012,Smith2010,Alan2014,jarus2014runtime,witkowski2013practical} and non-linear \cite{Fan2007,Lien2007,Qureshi2009,Tang2011,Sun2015PowerEfficientPF,Janacek2012,Guazzone2012,Zhang2013,mccullough2011evaluating,LUO2013,thesis,Li2016,Dhiman,canuto2016methodology,cupertino2015towards}.

Despite the increasing interest in the energy consumption issue of the data centers, little work has been done to systematically analyze and compare the performance of different software-based power consumption models.  These models in the literature are evaluated under different environment, experimental setup and analyzed using variants of formula to calculate the error.  To date, there have been relatively very few surveys conducted for server level software-based power consumption modeling.  Rivoire et al. proposed a power consumption model for servers and compared its performance with four other power models in a unified setup using a diverse set of applications \cite{Economou2006,Rivoire2008a}.  However, a key limitation of this work is that it fails to be comprehensive and only compares power models proposed before 2008.  M{\"o}bius et al. provided a comprehensive survey of different power models for predicting power or single-core or multi-core processors, virtual machines, and entire server \cite{mobius2013power}.  In addition, the work extracted the factors affecting the estimation error of the power models based on the literature review.  Dayarathna et al. in the year 2016, conducted a survey of different energy consumption modeling techniques covering more than 200 models \cite{Dayarathna2016}.  Though providing a detail literature review of different power models, the surveys \cite{mobius2013power,Dayarathna2016} lacks a comparative performance evaluation of the studied models.  Lin et al. in the year 2018, reported on the performance of different power models for disk, CPU and memory individually in a cloud system \cite{DBLP:journals/fgcs/LinWWWH18}.  However, the comparison does not involve the models for the overall server power consumption.  In this work, we conducted the evaluation of software-based power models in a unified setup.

\section{Conclusion}
The surging data center energy consumption with the rapid popularity of cloud services, big data analysis and IoT has led to crucial economic and environmental issues.   An increasing amount of research on power optimization for energy efficient designing and resource management has thus gained major attention in recent years.  Power modeling and prediction at different levels of data center plays a vital role in this context.  Many works on power modeling have been proposed in the literature aiming towards energy efficient computing.  Those models were evaluated using different experimental setup, benchmarking applications, power measurement technique and error calculation formula, which makes it difficult to compare their relative performance.  To our knowledge, this is the first work presenting a survey and comparative analysis of these models in a unified setup.

In this study, we present taxonomy and comparative evaluation of state-of-the-art software-based server power consumption models under a unified experimental setup.  For that purpose, we perform a series of experiments on three different server architectures.  The evaluation uses nine different tools and benchmarking applications having diverse resource utilization, for model development and evaluation.

Our experimental results show that among the single variable power models, interpolation has the least error while among the multi variable ones, SVM power model has the least error of estimation.  Comparing the overall performance for the different applications, the interpolation model gives the least error for CPU-intensive application, while SVM model gives the least error for CPU+memory, CPU+memory+disk, and CPU+memory+disk+network-intensive applications.  The lasso regression with 30 variables performs the worst with a maximum error of estimation for all the studied application types.   Our experiments reveal that the relative performance of these models remains on different server architectures.

When developing/using power consumption models in a computing environment, the following requirements should be considered.
\begin{enumerate}
\item \textit{Linear versus non-linear models:} The accuracy of the linear (regression) models mainly depends on the selection of features significantly related to power consumption which requires domain knowledge.  Moreover, linear models assume that the selected features have no correlation, which might not be the case.  On the other hand, the non-linear models such as SVM can capture the correlation between the features which results in less error of estimation.
\item \textit{CPU utilization dominance:} Most often the server's power is represented as a function of its CPU utilization, considering CPU to be dominant power consumer.  For the applications that are not CPU-intensive, this assumption breaks down.  It is advisable to consider at least memory utilization as it is the second most power consuming resource after CPU.  Memory utilization refers to accessing the Dynamic Random-Access Memory (DRAM) for requests which are not served by the three levels of cache (L1, L2 and L3).  The power consumption is directly related to the DRAM access through its controller.  However, when the memory used by an application is distributed across multiple memory controllers for better throughput, the DRAM accesses through controllers will increase.  This may lead to more power consumption as the number of accessed memory controllers increases.  Consequently, the impact of memory controllers should be considered in power models.
\item \textit{Server's idle power:} The idle power varies with the server architecture and assuming it to 70\% of its peak power, by one of the power models under study, may lead to drastically misleading predictions.  The more the server is energy efficient, the less is its idle power compared to the peak power.  But, an energy-aware scheduler might avoid placing the task on the energy efficient server predicting its energy consumption based on the assumption of idle power to be 70\% of its peak. 
\item \textit{Kernel Function:} The selection of the kernel for the SVM model should be done efficiently to yield most accurate results with least complexity.  The kernel function selection is dependent on the behavior of the training data set.
\item \textit{Quadratic and cubic utilization functions:} Combinations of linear, quadratic and cubic functions of different performance counters, selected using different variable selection models should be first used to select the variables representing the power consumption with high correlation.  The selected variables should be then used for model development.
\item \textit{Throughput versus performance counters:} The throughput-based power model has a better performance than the MVLR models.   However, the throughput-based requires calibrations of the regression coefficients for every application with a different throughput unit, for each server architecture type.  The MVLR models can be trained periodically for each server type.

\end{enumerate}

For Future research work, we propose investigations in the following directions.  First, we would like to investigate a multi-objective scheduling algorithm in conjunction with an energy model to optimize the energy consumption, performance and the quality of services in the data centers. Second, it would be valuable to compare the performance of software-based and hardware-based power models used in the literature of power modeling.

\begin{acks}
This work is supported by the Emirates Center for Energy and Environment Research of the United Arab Emirates University under Grant 31R101.  The authors would like to thank the anonymous reviewers for their valuable comments which helped us improve the content, quality, and presentation of this paper.
\end{acks}

%%\bibliographystyle{ACM-Reference-Format}
%%\bibliography{References}



\end{document}